\documentclass[runningheads,a4paper,12pt]{article}

\usepackage{cmap}
\usepackage{latexsym}
\usepackage[cp1251]{inputenc}
\usepackage[english]{babel}
\usepackage[unicode]{hyperref}
\usepackage{fancyhdr}
\usepackage{cmap}
\usepackage{amssymb}
\usepackage{tikz}
\usetikzlibrary{shapes,arrows, positioning}
\pgfmathtruncatemacro\distance{1}

\def\ral#1{\;\mathop{\longrightarrow}\limits^{#1}\;}

\def\blackbox{\;\vrule height 7pt width 7pt depth 0pt \;}

\def\arraystretch{1.2}

\def\be#1{\begin{equation}\label{#1}}
\def\ee{\end{equation}}

\def\re#1{(\ref{#1})}
\def\i{\item}

\def\bi{\begin{itemize}}
\def\ei{\end{itemize}}
\def\bn{\begin{enumerate}}
\def\en{\end{enumerate}}

\def\c#1{\left\{\begin{array}{lllll}#1\end{array}\right\}}

\def\by{\begin{array}{llllllllllllll}}
\def\ey{\end{array}}

\def\eam{\;\mathbin{{\mathop{=}\limits^{\mbox{\scriptsize def}}}}\;}

\def\ba{\left\{\begin{array}{llllllllll}}
\def\ea{\end{array} \right\}}
\def\bau{\left[\begin{array}{llllllllll}}
\def\eau{\end{array} \right]}
\def\beq{\begin{equation}}

\def\arraystretch{1.2}
\def\ra#1{\;\mathop{\to}\limits^{
\begin{picture}(0,5)
\put (-3,2){\makebox(0,1){$\scriptstyle #1$}}
\end{picture}
}\;}

\def\blackbox{\;\vrule height 7pt width 7pt depth 0pt \;}

\def\be#1{\begin{equation}\label{#1}}
\def\ee{\end{equation}}
\def\i{\item}

\def\re#1{(\ref{#1})}
\def\bn{\begin{enumerate}}
\def\en{\end{enumerate}}

\def\c#1{\left\{\begin{array}{lllll}#1\end{array}\right\}}

\def\by{\begin{array}{llllllllllllll}}
\def\ey{\end{array}}
\def\bi{\begin{itemize}}
\def\ei{\end{itemize}}
\def\bn{\begin{enumerate}}
\def\en{\end{enumerate}}
\def\eam{\mathbin{{\mathop{=}\limits^{\mbox{\scriptsize def}}}}}

\def\beenum{\begin{enumerate}}
\def\enenum{\end{enumerate}}
\def\ba{\left\{\begin{array}{llllllllll}}
\def\ea{\end{array} \right\}}
\def\bau{\left[\begin{array}{llllllllll}}
\def\eau{\end{array} \right]}
\def\beq{\begin{equation}}

\def\blackbox{\;\vrule height 7pt width 7pt depth 0pt \;}

\def\arraystretch{1.2}

\def\be#1{\begin{equation}\label{#1}}
\def\ee{\end{equation}}

\def\re#1{(\ref{#1})}
\def\i{\item}

\def\bi{\begin{itemize}}
\def\ei{\end{itemize}}
\def\bn{\begin{enumerate}}
\def\en{\end{enumerate}}

\def\c#1{
{\def\arraystretch{0.7}
\left\{\begin{array}{lllll}#1\end{array}\right\}}}

\def\by{\begin{array}{llllllllllllll}}
\def\ey{\end{array}}

\def\eam{\mathbin{{\mathop{=}\limits^{\mbox{\scriptsize def}}}}}

\def\ba{\left\{\begin{array}{llllllllll}}
\def\ea{\end{array} \right\}}
\def\bau{\left[\begin{array}{llllllllll}}
\def\eau{\end{array} \right]}
\def\beq{\begin{equation}}

\author{A. M. Mironov}

\title{New method of verifying cryptographic protocols 
based on the process model 
\footnote{This research has been financially supported by the Ministry of Digital Development, Communications and Mass Media of the Russian Federation and Russian Venture Company (Agreement No.004/20 dated 20.03.2020, IGK 0000000007119P190002)
}
}

\date{ 
Innopolis University, \\Leading reserch center\\
$\;$\\
amironov66@gmail.com
}

\begin{document}

\maketitle

\newcounter{theorem}
\newcounter{lemma}
\newcounter{fig}

\begin{abstract}

A {\bf cryptographic protocol (CP)}  
is a distributed algorithm  designed  to provide a
secure communication in an insecure environment.
CPs are used, for example, in electronic payments, electronic voting procedures, database access systems, etc. 
Errors in the CPs can lead to great financial and social damage, therefore it is necessary to use mathematical methods to justify the correctness and safety of the CPs.
 In this paper, a new mathematical model of a CP is introduced, which allows one to describe both the CPs and their properties. It is shown how, on the basis of this model, it is possible to solve the problems of verification of CPs.
\end{abstract}

\section{Introduction}

\subsection{A concept of a cryptographic protocol}

A {\bf cryptographic protocol (CP)} 
is a distributed algorithm that describes the order in which messages are exchanged between agents. Examples of such agents are computer systems, bank cards, people, etc.

To ensure  security properties of a CP (such as, for example, the secrecy of transmitted data), cryptographic transformations (encryption, electronic signature, hash functions, etc.) can be used in the CP.

We assume that the cryptographic transformations used in CPs are ideal, i.e. satisfy some axioms expressing, for example, the impossibility of extracting plain texts from encrypted texts
without knowing  of the corresponding cryptographic keys.

\subsection{Vulnerabilities in cryptographic protocols}

Many CP vulnerabilities are related  not with poor cryptographic qualities of the cryptographic primitives used in them, but with logical errors in protocols. For example, a logical error was found in the CP for logging into a Google portal that allows a user to identify himself only once and then get access to various applications (such as Gmail or Google Calendar), allowing a dishonest service provider to impersonate any of its users. 

There are many other examples of CPs (see for example \cite{1}-\cite {15}), which have been used for a long time in security-critical systems, but then it was discovered that these CPs contain vulnerabilities of the following type:
\bi
\i participants of these CPs can receive distorted messages (or even lose them) as a result of interception, deleting or distorting of transmitted messages by the adversary, which violates the integrity property,
\i the adversary can discover a secret information 
contained in the intercepted messages as a result of erroneous or malicious actions of CP participants.
\ei

Vulnerabilities were also detected in one of the most well-known CPs {\bf Kerberos} \cite{kerbprot}.
The absence of vulnerabilities in the patched version of Kerberos was justified in \cite{kerberos}.
There are many other examples of CP vulnerabilities used to 
authentication for cell phone providers, ATM cash withdrawals, e-passports, electronic elections, etc. 

All of the above examples justify the fact that 
an informal analysis of the required properties 
is not enough for CPs used in the security critical systems, 
it is necessary
\bi
\i to construct a {\bf mathematical model} of the analyzed CPs,
\i describe properties of analyzed CPs in the form of a mathematical objects called {\bf specifications} of these CPs, and
\i to construct proofs of statements that the analyzed CPs meet (or do not meet) the specifications, the procedure for constructing such proofs is called {\bf verification} of the analyzed CPs.
\ei

In this work, a new mathematical model of CPs is constructed.
In terms of this model it is possible to express such properties of correctness of CPs as, for example, integrity of transmitted messages (i.e., justification of the following property of the analyzed CPs: messages sent by one participant of a CP to another participant of this CP, reach the recipient in an undistorted form).

\subsection{Historical overview of methods for verifying cryptographic protocols}

Historically, first formal approach for CP verifying
was the BAN-logic of  Burrows M., Abadi M., and Needham R., 
\cite{5}. 
This approach has very large limitations, in particular, it does not allow considering the case of unlimited generation of sessions of the analyzed protocol.

A more popular approach to CP verification is the strand spaces formalism developed by Joshua D. Guttman, Jonathan C. Herzog, F. Javier Thayer Fabrega, \cite{strand1}-\cite{strandlast}. 
Among the works devoted to the description of various formalisms designed for modeling and verification of CPs, it should also be noted articles 
\cite{hartog}-\cite{Security}.

One of the CP verification formalisms is the approach associated with the use of Horn clauses and Constraint Systems, developed in the works of Abadi, Blanchet, Cortier and other specialists \cite{verc}. Among other CP models, 
the most popular ones are logic models
(see for example \cite{5}, \cite{11}, \cite{27}). These models make it possible to reduce the problems of CP verification to the problems of constructing proofs of theorems that CPs under analysis meet their specifications. Algebraic and logical approaches to CP verification are also considered in \cite{rsm} - \cite{abf}.

\section{Sequential and distributed processes}

In this paper, we outline the concepts of sequential and distributed processes. These concepts are basic
mathematical objects for building a CP process model. 
This model is a development of the {\bf Calculus of Cryptographic Protocols} of Abadi-Gordon
({\bf SPI-calculus}, \cite{82}). It can serve as a theoretical basis for a new method for verifications of CPs, where CP verification means the construction of a mathematical proof that an analyzed CP has the desired properties. Examples of such properties are {\bf integrity} and {\bf secrecy} properties. In the process model described in this text, CPs and their formal specifications are represented as distributed processes. 

One of the most important advantages of the proposed CP process model is the low complexity of proofs of CP correctness. In particular, this model eliminates the need to build the sets of all reachable states of the analyzed CPs. This provides an important advantage when analyzing sets of states of the analyzed CP in the case when  sets of these states are potentially unlimited. Another important advantage of the proposed CP model is the high degree of automation of solving the CP verification problem based on this model.

\subsection{Auxiliary concepts}

\subsubsection{Types, constants, variables, function symbols}

We assume that there are given sets 
$Types$, $Con$, ${\it Var}$ and $Fun$.
The elements of these sets are called
{\bf types},
{\bf constants},
{\bf variables}, and
{\bf function symbols (FS)}, respectively.
Each element
$x$  of $Con$, ${\it Var}$ and
$Fun$  is associates with 
some type 
 $\tau (x) \in Types$, 
 and if $x \in Fun$, then $\tau (x )$ has the form $(\tau_1,\ldots,\tau_n)\to \tau$,
 where 
$\tau_1,\ldots, \tau_n, \tau \in Types.$

\subsubsection{Terms}

The concept of a {\bf term} is defined inductively. 
Each term $e$ is associated with a type 
 $\tau(e) \in Types$.
The definition of a term is as follows:
  \bi
  \i $\forall\,x\in  Con\cup {\it Var}\;\;x$
  is a term of the type
  $\tau(x)$,
\i if $f\in Fun$, $e_1,\ldots, e_n$ are terms, and
$\tau(f)$ has the  form 
$$(\tau(e_1),\ldots, \tau(e_n))\to \tau,$$
then the notation $f(e_1,\ldots, e_n)$
is the term of the type  $\tau$.
\ei

We will use the following notations:

\bi
  \i   $Tm$ denotes the set of all terms,
\i     $\forall\,e\in Tm\;\;
     {\it Var}_{e}$ denotes the set of all variables,
occurred in $e$,
\i $\forall\,X\subseteq  {\it Var}\;\;
Tm(X)$  
denotes the set $\{e\in Tm\mid
{\it Var}_e\subseteq X\}$,
\i $\forall\,E\subseteq Tm,
\forall\,\tau\in Types
\; E^\tau$
denotes the set $\{e\in E\mid \tau(e)=\tau\}$.
\ei

Let $e,e'\in Tm$.
The term $e$ is called a {\bf subterm} of the term $e'$,
if  either $e=e'$, or $e'$ has the form 
$f(e_1,\ldots, e_n)$, where $f\in Fun$,
and $\exists\,i\in\{1,\ldots, n\}$: $e$ is a subterm of the term
$e_i$. 

The notation $e\subseteq e'$, where $e,e'\in Tm$,
means that $e$ is a subterm of  $e'$.

Below, for each considered function of the form
$\varphi:E\to E'$, where $E,E'\subseteq Tm$, 
we will assume that 
$\forall\,e\in E\;\;
\tau(\varphi(e))=\tau(e)$.

\subsubsection{Examples of types}

We shall assume that  $Types$ has the following types:
\bi
\i type {\bf A}, terms of this type are called 
{\bf agents},
\i type {\bf C}, terms of this type are called 
{\bf channels}, they denote  communication channels 
used by agents for communication
with each other by sending messages,

\i type {\bf K}, terms of this type are called {\bf keys}, they denote cryptographic keys,
that agents can use to encrypt or decrypt messages,
\i type {\bf M}, terms of this type are called {\bf messages}, they denote
messages, that agents can send to each other in the work flow,
\i  type {\bf P}, terms of this type are called  {\bf processes}.
\ei

The notations $Agents$, $Channels$, $Keys$ 
and $Processes$
denote the sets of all agents,
channels, keys, and processes, respectively.

We will use the following conventions and notations:
\bi
\i  $Channels$ has a 
constant denoted by $\circ$, and
called an {\bf open channel},
\i an occurrence of a key $k$ in a term $e$ is said to be {\bf hidden},
if this occurrence is first occurrence 
of $k$ in a subterm of the form $k(e')\subseteq e$,
\i $\forall\, A\in {\it Var}^{\bf A}$ the
set ${\it Var}$ has the variable 
$A^-\in {\it Var}^{\bf K}$,
called the {\bf private  key} of  agent $A$,
\i
type ${\bf M}$ includes 
 any other types from $Types$, i.e. a
term of any type
is also a term of  type ${\bf M}$,\i 
$\forall\,n\geq 1$ set
 $Types$ has  type ${\bf M}_n$, 
whose  values are tuples of length $n$,
consisting of values of type ${\bf M} $,
\i  set $Var$  contains {\bf shared variables},
each such variable has the form 
$x_{P_1\ldots P_n}$, where $P_1,\ldots, P_n$ are 
different  constants of the type ${\bf P}$.
\ei

\subsubsection{Examples of function symbols}

We will assume that $Fun$ contains the following FSs.
\bi
\i FS $tuple_n$, where $n\geq 1$ and
$\tau(tuple_n)=(\;\underbrace{{\bf M}, \ldots, {\bf M}}_n\;)\to {\bf M}_n$.

For each list $(e_1,\ldots, e_n)$  of terms 
the term
$tuple_n(e_1,\ldots, e_n)$ 
will be denoted by a shorter notation
 $(e_1,\ldots, e_n)$.
 
\i FS $pr_{n,i}$, where  $n\geq 1$, 
$i\in\{1,\ldots, n\}$, and
$\tau(pr_{n,i}) = {\bf M}_n\to {\bf M}$.

$\forall\,e\in Tm^{{\bf M}_n}$ the term
$pr_{n,i}(e)$  is the $i$--th component of the tuple
$e$, 
this term will be denoted by the notation $(e)_i$.

\i FS $h$ (possibly with indices)  of type
${\bf M}\to {\bf M}$ type.

The term $h(e)$ denotes the {\bf hash function}
 value of the message $e$.

\i FSs $encrypt$ and $decrypt$ 
of  type
$({\bf K}, {\bf M})\to {\bf M}$.

Terms of the form $encrypt (k, e) $ and $decrypt (k, e) $ 
denote messages received by encrypting (and decrypting, respectively) the message $e$ on the key $k$.

\i FS $public\_key$ of type
${\bf A}\to {\bf K}$.

Term  of the form
$public\_key(A)$ 
is called the
 {\bf public key} of agent $A$.

Terms of the form  $encrypt(k,e)$ and 
$encrypt(public\_key(A),e)$
will be denoted by the notations
$k(e)$ and $A(e)$ respectively, 
this terms are called
{\bf encrypted messages}.

\i FS  $dig\_signature$ of  type
$({\bf M} , {\bf A})\to {\bf M}$.

A term of the form
$dig\_signature(e,A)$ denotes a {\bf digital signature} of the message $e$, made by agent $A$.

The triple
$(e,A,dig\_signature(e,A))$ will be denoted by
$(e)_A$.
\ei

\subsubsection{Expressions}

An {\bf expression} is a notation of one of the following forms:
\bi
\i any set of terms $E\subseteq Tm$,
\i  $X_P$, where $P\in Processes$,
\i $M_c$, where $c\in Channels$, 
\i  $k^{-1}(E)$, where $k\in Keys$, and $E$ is an expression,
\i $E\cap E'$,
$E\cup E'$,
$\neg E$, where 
$E,E'$ are expressions.
\ei

The set of all expressions is denoted by $Expr$. 
$\forall\,E\in Expr$ the notation ${\it Var}_{E}$
denotes the set of all variables occurred in $E$.

If $E=\{e\}$, where $e\in Tm$, then such an expression will be denoted without brackets.

\subsubsection{Formulas}

An {\bf elementary formula (EF)} is a notation of one of the following forms:
\bi
\i 
$E= E'$, 
$E\subseteq  E'$, 
$E\supseteq E'$, 
where $E,E'\in Expr$,
\i $x\,\bot\,P$, $x\,\bot\,C$, 
where $x\in
 {\it Var}$, $P\in Processes$,
 $C\subseteq  Channels$,
\i $k\,\bot_{\bf K}\,P$, $k\,\bot_{\bf K}\,C$, 
where $k\in
 {\it Keys}$, $P\in Processes$,
 $C\subseteq  Channels$.
\ei

Examples of EFs:
\be{sdfsadfgsdfgsdfd}\left.\by
decrypt(k,k(e))=e,&\mbox{where } k\in {\it Var}^{\bf K}, e\in
Tm\\
decrypt(A^-,A(e))=e,&\mbox{where } A\in {\it Var}^{\bf A}, e\in
Tm\\
pr_{n,i}(e_1,\ldots,e_n)=e_i,&
\mbox{where } n>0,i\in \{1,\ldots,n\},\\&
e_1,\ldots,e_n\in Tm.
\ey\right\}\ee

A {\bf formula} is a  set of EFs.
The set of  all formulas is denoted  by the notation $Fm$.
$\forall\,\beta\in Fm$ the notation
${\it Var}_\beta$ denotes the set of all variables,
occurred in $\beta$. 

Each formula $\beta\in Fm$ 
defines a congruence $\sim_\beta$
on $Fun$ --algebra $Tm$:
$\sim_\beta$ is an intersection 
of all congruences $\sim$  
on  $Tm$ satisfying the condition: 
$\forall\,(e=e')\in \beta\quad
e\sim e'.$

Below, the equality of terms is understood up to the congruence $\sim_\beta$, where $\beta$ consists of EFs whose form coincides with one of the forms  in \re{sdfsadfgsdfgsdfd}.

\subsubsection{Bindings}

A {\bf binding} is a function of the form 
$\theta:{\it Var} \to Tm$.

We say that a  binding $\theta$ binds the variable 
$x \in {\it Var}$ with the term $ \theta (x) $.

We will use the following notations:
\bi\i
the set of all bindings is denoted by the symbol
 $\Theta$,
\i $id$ denotes identical binding:
$\forall\,x\in {\it Var}\;\;id(x)=x$,
\i $\forall\,X\subseteq {\it Var}$ notation $\Theta_X$ denotes 
the set $$\{\theta\in \Theta\mid \forall\,x\in {\it Var} \setminus X\;\;\theta(x)=x\},$$
\i a
binding $\theta\in \Theta$ 
 can be denoted by the notations 
$$x\mapsto\theta(x)\quad \mbox{or}\quad
(\theta(x_1)/x_1,\ldots, \theta(x_n)/x_n)$$
(second notation is used when 
$\theta\in \Theta_{\{x_1,\ldots, x_n\}}$),
\i 
 $\forall\,\theta\in \Theta,\;\forall\,
e\in Tm$ 
the notation $e^{\theta}$
denotes a term derived from $e$
by replacing $\forall\,x\in {\it Var}_e$
each occurrence  of 
  $x$ in $e$ by the term $\theta(x)$,
\i 
 $\forall\,\theta\in \Theta,\;\forall\,
E\subseteq Tm$ 
the notation $E^{\theta}$
denotes the set $\{e^\theta\mid e\in E\}$, 
\i
$\forall\,\theta,\theta'\in \Theta$ the
notation $\theta\theta'$ denotes the binding 
$x\mapsto (x^{\theta})^{\theta'}$.
\ei

\subsection{Sequential processes}

\subsubsection{Actions}

An {\bf action} is a notation of one of the following forms:
$$\by
c!e,\quad c?e,\quad e:=e', \quad
\mbox{where } c\in Channels,\;e,e'\in Tm,\ey$$
which are called 
{\bf sending} message $e$ to  channel $c$, 
 {\bf receiving} message $e$ from channel $c$, 
and  {\bf assignment},
respectively. 

Actions of the form 
$c!e$ and $c?e$ are called {\bf external} actions, 
and actions of the form $e:=e'$ are called {\bf internal}
actions.

The set of all actions is denoted by the notation $Act$. 
$\forall\,\alpha\in {Act}$ 
the set of all variables occurred in $\alpha$, 
is denoted by the notation ${\it Var}_{\alpha}$.

If $\theta\in \Theta$ and $\alpha\in Act$, then the notation
$\alpha^{\theta}$ denotes an action
$c^{\theta}!e^{\theta}$,
$c^{\theta}?e^{\theta}$ and
$e^{\theta}:=(e')^{\theta}$, if $\alpha=$
$c!e$,
$c?e$ and
$e:=e'$, respectively.

In some cases, to facilitate a perception, actions can be written in brackets, i.e., for example, instead of $c! e$, the notation $(c! e)$ might be used, etc.

\subsubsection{A concept of a sequential process}

A {\bf  sequential process (SP)} is a triple $(P,X, \bar X)$,
whose components have the following meaning:
\bi \i $P$ is a graph with a selected node
(called an {\bf initial} node, and denoted by $P^0$), 
each edge of which is labeled by an action
$\alpha \in Act$, 
\i $X \subseteq {\it Var} \cup Con $ is a set of
{\bf initialized variables} and constants,  $\circ \in X$,
\i $\bar X \subseteq X \cap Var$ is a set of {\bf hidden variables}, these variables denote secret keys,
hidden channels, and objects with unique values called {\bf nonces}.
 \ei

A SP is a formal description of the behavior of a dynamic system, which works by sequentially performing actions related to sending/receiving messages and initializing uninitialized variables.
 
For each SP $(P,X,\bar X)$
\bi
\i  this SP can be abbreviated by the same symbol $P$ as the corresponding graph, the set of nodes of the graph $P$
also is denoted by $P$,
\i nodes of graph $P$, which have no outgoing edges, 
are said to be {\bf terminal} and are denoted by $\otimes$,
\i notations $X_P$,  $\bar X_P$
denote the corresponding components of the SP $P$,
\i 
${\it Var}_P$ denotes the set of all variables occurred in  $P$,
 \i if $P$ has no edges and  $X_P=\emptyset$, then 
 $P$ is denoted by {\bf 0}.
\ei

 Each SP is associated with a constant from 
$Processes$, called a {\bf name} of this process. 
 In order to simplify notations, we will denote the names of processes with the same notations 
 that denote the processes themselves.

Actions of the form 
$\circ ! e$ and $\circ ?e$ will be shortened as
$ ! e$ and$ ?e$ respectively.

\subsubsection{Adversary process}

The {\bf adversary process} is a SP 
$P_*$ with the following features:
\bi
\i the SP graph $P_*$ has a single node,
\i $Con\subseteq X_{P_*}$,
 $\forall\,\tau\in Types$ the sets $\bar X_{P_*}$
and $X_{P_*}\setminus \bar X_{P_*}$
have a countable set of variables of the type $\tau$,
\i $\forall\,\alpha\in Act$  graph ${P_*}$ has an edge
labeled by $\alpha$.
\ei

Below we assume that $P_*$ is the only SP
under consideration, whose graph has cycles.

\subsubsection{States of sequential processes}

Let $P$ be a SP.
A {\bf state} of $P$ is a 4-tuple 
$s=(v,\alpha,X,\theta)$,  where 
\bi\i $v\in P$ is a {\bf current node},
\i $\alpha\in \{init\}\sqcup Act$ is a {\bf current action},
\i  $X\subseteq  {\it Var}$ is a {\bf current set of  initialized variables}, and
\i $\theta\in \Theta$ is a {\bf current binding}.
\ei

Components of  $s$ are denoted by $v_s$, 
$\alpha_s$,
$X_s$, and
 $\theta_s$,  respectively.

A state of the SP $P$ is said to be 
 {\bf initial}, and is denoted by 
$\odot$, if it has the form 
$(P^0, 
init,
X_P,
id).$

\subsubsection{An execution of a sequential process}
\label{hfdsklds}

Let $P$ be a SP. An {\bf execution} of $P$ can be understood 
as a walk through the graph $P$, starting from  $P^0$, 
with the execution of actions that are labels of traversed edges.

Each step of an execution of $P$ is associated with 
\bi\i a state of $P$, 
called a {\bf  current state} at this step
(a current state at first step is $\odot$),
and
\i a {\bf current channels state}, which is a family of sets
$$M=
\{M_c\subseteq Tm\mid c\in Channels\}.$$
\ei

If a current step of the execution of $P$ is not a final step, 
then  the following actions are performed at this step:
\bi \i 
the current state $s$ on this step
is changed  on a state $s'$, which will be a current state at the next step of the execution: 
if $s$ has the form $ (v, \alpha, X, \theta) $, 
then there is selected
an edge of $P$ outgoing from $v$, 
whose label $\alpha'$ meets one of the folllowing conditions:
\be{asdfsafgewe4444}
\!\!\!\!\!\!\!\!\!\!\!
\left.\by
({\rm a}) & \alpha'=c!e,\; 
c^{\theta}\in X^{\theta}$, $e\in Tm(X)\\
({\rm b})&\alpha'=c?e,\; 
c^{\theta}\in X^{\theta},\; \exists\,\hat \theta\in \Theta_{{\it Var}\setminus X}:
(e^{\hat \theta})^{\theta}\in M_{c^{\theta}}\\
({\rm c}) & \alpha'=(e:=e'),\;  e'\in Tm(X),\;
\exists\,\hat \theta\in \Theta_{{\it Var}\setminus X}:
(e^{\hat \theta})^{\theta}=
(e')^{\theta}
\ey\right\}
\ee
and components  of  $s' = (v', \alpha', X', \theta')$ 
have the following form:
 $v'$ is the end of the selected edge, 
 $\alpha'$ is the label of the selected edge,
and

\bi
\i if (a) in \re{asdfsafgewe4444} holds, then $X'= X$,
$\theta'=\theta$, 
\i if (b) or (c) in \re {asdfsafgewe4444} holds,
 then $X'=X\cup {\it Var}_e$, 
$\theta'=\hat \theta\theta$, and
 \ei
\i a replacement of the current channels state 
$M$ with the channels state $M'$, 
which will be the current channels state at the next step of the execution: $M'$ either is equal to $M$, or is obtained by adding terms to the sets from $M$, and
\bi \i
this adding can be performed by  $P$ as well as  those SPs
that use shared channels with $P$, and 
\i if (a) in \re {asdfsafgewe4444} holds, 
then one of such addings is that $P$ adds 
the term $e^{\theta}$ to the set 
$M_{c ^{\theta}}$. \ei \ei

We will say that $s'$ is obtained by a {\bf transition} from $s$, and denote this by the notation $s\to s'$.

During each execution of each SP $P$ the variables from 
${\it Var}_{P}$ have the following features:
$\forall\,x\in {\it Var}_{P}$
\bn\i
if $x\not\in X_{P}$, then at the initial step of each execution of  $P$ the variable $x$ is not initialized, i.e. there is no value associated with $x$, \i\label{sdafasgsdfgfds} 
if $x\in \bar X_{P}$ and $x$ is not a shared variable,
then 
at first step of each execution $Exec$ of $P$ this variable is associated with a {\bf unique value}, i.e. such 
a value that differs from 
values associated with other initialized variables at $Exec$,
and  from values associated with initialized variables at any
execution $Exec'\neq Exec$ of any SP, 
\i if a variable from $\bar X_{P}$
 is shared and has the form $x_{P_1\ldots P_n}$, then \bi\i
  $P_1,\ldots, P_n$ is a list of names of all SPs, 
  executed together with $P$ (and $P$ is one 
  of the SPs in this list), which have the variable
  $x_{P_1\ldots P_n}$ among his hidden variables, 
  and \i at the initial moment of each joint execution of SPs from the list  $P_1,\ldots, P_n$ variable $x_{P_1\ldots P_n}$  is initialized in all these SPs with the same value, which is unique, i.e. has the properties described in the point \ref{sdafasgsdfgfds}.  
 \ei
\en

\subsection{Operations on sequential processes}

\subsubsection{Prefix action}

A {\bf refined action} is a triple 
$\tilde \alpha=(\alpha, \hat X, \bar X)$, where 
$\alpha\in Act$, and $\hat X, \bar X$
are disjoint subsets of the set  ${\it Var}_{\alpha}$.

We will denote the refined action
$\tilde \alpha=(\alpha, \hat X, \bar X)$ by the
notation obtained from the notation of the action $\alpha$
by replacing  each variable $x\in {\it Var}_\alpha$ 
to  $\hat x$ or $\bar x$, 
if $x\in \hat X$ 
or 
$x\in \bar X$, respectively.

Let  $\tilde \alpha=(\alpha, \hat X, \bar X)$
be a refined action and  $P$ be a SP.
An operation of a {\bf prefix action} maps
the pair $(\tilde \alpha, P)$
to a SP $\tilde \alpha.P$, 
having the following components:
\bi
\i a graph of the SP $\tilde \alpha.P$ is obtained 
by adding \bi\i a new node $v$ to $P$, which will be an initial node in $\tilde \alpha.P$, 
and 
\i an edge $v\ra{\alpha}P^0$,
\ei
\i $X_{\tilde \alpha.P}=(X_P\cup {\it Var}_\alpha)\setminus 
\hat X,\quad 
\bar X_{\tilde \alpha.P} = \bar X_P\cup  \bar X$.
\ei

Below we will omit the symbol $\sim$ in the notations of the refined actions.

\subsubsection{Choice}

Let $P_I=\{P_i\mid i\in I\}$
be a family of SPs.

The notation $\sum_{i\in I}P_i$ denotes a 
SP $(P,X,\bar X)$, called a
{\bf choice} from $P_I$.
Its components are defined as follows:
\bi
\i the graph $P$ is obtained by adding 
 to the union of disjoint copies of graphs from 
${P_I}$ 
\bi\i  a new node 
$P^0$, which will be the initial one
in $P$, and
\i edges $P^0\ra{\alpha}v$, 
corresponding to  edges of the form
$P_i^0\ra{\alpha}v$,\ei
\i $X$ and $\bar X$ are unions 
of the corresponding components of SPs from
$P_I$.
\ei

If the set of indices $I$ has the form 
$\{1,\ldots, n\}$, then SP $\sum_{i\in I}P_i$
can also be denoted  by $P_1+\ldots+P_n$.

\subsubsection{Renaming}

A {renaming} is a partial injective function 
$\zeta:Var\to Var$, where for each shared variable 
$x_{P_1\ldots P_n} \in Dom(\zeta)$
the variable $\zeta(x_{P_1\ldots P_n})$
has the form $y_{P_1\ldots P_n}$. 

For each renaming $\zeta$,
each term $e$ and each SP $P$
the notations $e^{\zeta}$ and $P^\zeta$
denote a term or a SP respectively, 
obtained from $e$ or $P$
by replacing $\forall\,x\in Dom(\zeta)$ of
each occurrence of $x$ by $\zeta(x)$.

If ${\it Var}_P\subseteq Dom(\zeta)$, then the 
SPs $P$ and $P^{\zeta}$ 
are assumed to be the same.

\subsection{Distributed processes}

\subsubsection{A concept of a distributed process}

Let $P_I=\{P_i\mid i\in I\}$ be a family of SPs. 

$\forall\,i\in I$ let  $\tilde X_{P_i}$ be a set of variables
 from ${\it Var}_{P_i}$, which
 either do not belong to $X_{P_i}$,
 or belong to $\bar X_{P_i}$ and are not shared.

We shall assume that for each family of SPs $P_I$ 
under consideration the sets $\tilde X_{P_i}$ are disjoint 
(if this is not the case, then we rename accordingly
variables in SPs from the family $P_I$). 

A {\bf distributed process (DP)}
corresponding to the  family $P_I$
is an object denoted by the notation
$\prod_{i\in I}P_i$. 
A DP is a model of a distributed algorithm, 
components of which are SPs from the family $P_I$, 
interacting by transmitting  messages through channels. 
The meaning of a DP concept  is explained in section
\ref{vypraspr}.

If $P$ is a DP of the form $\prod_{i\in I}P_i$, then 
\bi
\i ${\it Var}_P=\bigcup_{i\in I}{\it Var}_{P_i}$,
  $X_P=\bigcup_{i\in I}X_{P_i}$,
$\bar X_P=\bigcup_{i\in I}\bar X_{P_i}$,

\i if $\zeta$ is a renaming,  then\bi\i
the notation 
$P^\zeta$ denotes the DP $\prod_{i\in I}P_i^\zeta$,
\i if  ${\it Var}_P\subseteq Dom(\zeta)$, 
then  $P$ and $P^{\zeta}$ 
are assumed to be the same,\ei
\i $P$ can be denoted by the notation
\bi\i
$(P_1,\ldots, P_n)$,  
if $I=\{1,\ldots, n\}$, or\i 
$Q^\infty$, if $I$ is 
a set of natural numbers, and all SPs in the family
$P_I$  coincide with the  SP $Q$.
\ei
\ei

If  $P_I=\{P_i\mid i\in I\}$ is  a family of DPs, 
and each DP $P_i$ in $P_I$ corresponds to a family of SPs 
$\{Q_{i'}\mid {i'\in I_i}\}$, 
where the sets $I_i\;(i\in I)$ are disjoint 
(if this is not the case, then
we will replace these sets with appropriate disjunctive copies),
then the notation $\prod_{i\in I}P_i$ denotes
a DP corresponding to the  family of SPs
$\{Q_i\mid i\in \bigsqcup_{i\in I}{I_i}\}.$

If DP $P$ has the form 
$\prod_{i\in I}P_i$, then the notation $P^*$ 
denotes the DP $\prod_{i\in I\sqcup\{*\}}P_i$,
where $P_*$ is the adversary process.

\subsubsection{A concept of a state of a
distributed process}

Let $P$ be a DP of the form $\prod_{i\in I}P_i$.

A {\bf state} of $P$ is a pair $S$ of the following objects:
\bi
\i a set $\{s^S_{P_i}\mid i\in I\}$ of
states of SPs from $P_I$,
\i a {\bf channel state}: $M^S=
\{M^S_c\subseteq Tm\mid c\in Channels\}$.
\ei

A state $S$ of DP $P$ is said to be {\bf initial},
and is denoted by $\odot$, if 
$$
\forall\, i\in I\;\;s^S_{P_i}=\odot,\quad
\forall\,c\in Channels\;\;M^S_c=\emptyset.$$

If $S$ is a  state of the DP $P=\prod_{i\in I}P_i$,
and $i\in I$, then
\bi\i 
notations $v_{P_i}^S$, $\alpha_{P_i}^S$, 
$X_{P_i}^S$,  $\theta_{P_i}^S$ denote the corresponding 
components of the state $s^S_{P_i}$,
\i notation $V^S$ denotes the set
$\{v_{P_i}^S\mid i\in I\}$,
\i notation $\theta^S$ denotes a binding, such that
$$\forall\,i\in I,\;
\forall\,x\in X_{P_i}^S\quad
\theta^S(x)=\theta^S_{P_i}(x).$$
\ei

\subsubsection{An execution of a distributed process}
\label{vypraspr}

Let $P$ be a DP of the form $\prod_{i\in I}P_i$.

An {\bf execution} of $P$ 
can be understood as non-deterministic interliving 
of executions of SPs from $P_I$. 
At each step of an execution of  $P$
\bi
\i at most one SP from $P_I$ performs its current action, and 
\i other SPs from $P_I$ are in the waiting status. 
\ei

An execution of a DP $P$ can be formally defined as a
generation of a sequence of states of this
DP (starting with $\odot$),  in which each state $S$ that is not terminal, is associated with the next state
$S'$ 
by a  {\bf transition relation}, which means the following:
$\exists\,i\in I$:
\be{sdfagr33gr356ujyhfgd}\!\!\!
\by
s^S_{P_i}\to s^{S'}_{P_i}, \quad
\forall\,i'\in I\setminus\{i\}\quad
s^{S'}_{P_{i'}}=s^{S}_{P_{i'}},
\quad
\mbox{and if $s^{S'}_{P_i}=(v,\alpha,X,\theta)$, then}
\\
\mbox{if
$\alpha=c!e$, then }
\left\{
\by
 M^{S'}_{c^{\theta}}=
 M^{S}_{c^{\theta}}\cup\{e^{\theta}\},\\
 M^{S'}_{c'}= M^{S}_{c'} \mbox{ when $c'\neq c^{\theta}$}
 \ey\right\},
 \mbox{otherwise $M^{S'}= M^{S}$.}\ey\ee

For each states $S,S'$ of DP $P$
\bi\i 
 $S\to S'$ means that 
$S$ is related with $S'$ by a transition relation, 
\i $S\ral{\!\alpha_{P_i}}S'$ means that 
$S\to S'$, and \re{sdfagr33gr356ujyhfgd} holds,
\i
 $S\Rightarrow S'$ means that either $S=S'$,
or there is a sequence
$S_0,\ldots, S_n$  of states of  $P$, 
such that $$S_0=S,\quad
S_n=S', \quad
\forall\,i=0,\ldots, n-1\quad S_{i}\to S_{i+1}.$$\ei

A state $S$ of $P$ is said to be {\bf reachable}, 
if $\odot\Rightarrow S$.

The set of reachable   states of $P$ is 
denoted by $\Sigma_P$.

\subsection{Schemes of distributed processes}
\label{fadsgadsg3w54ggfd}

\subsubsection{A concept 
of a scheme of a distributed process}

Let $P$ be a DP of the form
$\prod_{i\in I}$, and $\forall\,i\in I$ 
SP $P_i$ has the form 
\be{dfasfafggrea442gfdgfad}
 \alpha_{1}.\;\ldots \; \alpha_{n}.P'_i.\ee

The sequence of actions  $\alpha_{1} \ldots  \alpha_{n}$
and SP $P'_i$ will be called a {\bf prefix} and
a {\bf postfix} of SP $P_i$, respectively.

If \bi\i each external actions in the prefix of
$P_i$ is a sending (receiving) a message to (from) 
a certain SP $P_j\in P_I$, and \i the action of 
SP $P_j$  corresponding to the receiving (sending) this
 message is in the prefix of $P_j$, \ei
 then these dependencies between actions can be expressed 
 as a {\bf scheme} of  DP $P$, which has the following form:
\bi
\i each  SP $P_i\in P_I$ is represented in this scheme by a 
{\bf thread}, i.e. by a vertical line, on which 
there are marked 
points corresponding to  nodes of the graph $P_i$ 
belonging to the prefix  of $P_i$ 
(the upper point of the thread corresponds to $P^0_i$), and 
\bi 
\i near each such point it might be specified an identifier of the corresponding node, 
\i near the upper point of the thread a name of SP $P_i$ is specified,   
\i if $P'_i\neq {\bf 0}$, 
then the postfix name $P'_i$ is specified near the bottom point of the thread,
\i the segments connecting the neighboring points of the thread correspond to  edges of $P_i$ related to the prefix of $P_i$, 
 there are the specified labels of the corresponding edges
 beside these segments,
\ei
\i for each segment $O$ of the thread 
connecting neighboring points,
 if the corresponding action is sending a message, 
 then there is an arrow in the scheme, such that 
\bi\i the start of this arrow lies on the segment  $O$, and
\i the end of this arrow lies on the segment  $O'$, 
the label of which is an action 
of the corresponding SP $P_j\in P_I$ to receive this message.
\ei
\ei

For example if $P_i= \alpha_{1}.\;\ldots\;  \alpha_{n}.P'_i$, where $\alpha_{1}$ is a sending,
and $\alpha_{n}$ is a receiving, and $A^0,\ldots, A^n$ are identifiers of the corresponding nodes of  $P_i$,
then a thread corresponding to 
$P_i$ has the following form: 
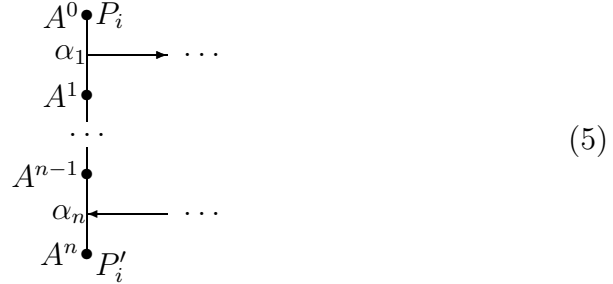
\be{afdsgasd342t34t5ytwe}\by
\begin{picture}(0,90)
\put(0,90){\circle*{4}}
\put(0,60){\circle*{4}}
\put(0,30){\circle*{4}}
\put(0,0){\circle*{4}}
\put(0,90){\line(0,-1){40}}
\put(0,0){\line(0,1){40}}
\put(0,45){\makebox(0,0){$\ldots$}}
\put(0,75){\makebox(0,0)[r]{$ \alpha_{1}$}}
\put(0,15){\makebox(0,0)[r]{$ \alpha_{n}$}}
\put(-3,90){\makebox(0,0)[r]{$A^0$}}
\put(-3,60){\makebox(0,0)[r]{$A^1$}}
\put(-3,30){\makebox(0,0)[r]{$A^{n-1}$}}
\put(-3,0){\makebox(0,0)[r]{$A^n$}}
\put(3,-5){\makebox(0,0)[l]{$P_i'$}}
\put(0,75){\vector(1,0){30}}
\put(3,90){\makebox(0,0)[l]{$P_i$}}
\put(0,75){\vector(1,0){30}}
\put(30,15){\vector(-1,0){30}}
\put(36,75){\makebox(0,0)[l]{$\ldots$}}
\put(36,15){\makebox(0,0)[l]{$\ldots$}}
\end{picture}\ey
\ee

\subsubsection{Examples of schemes of distributed processes}
\label{pkergnmf43564654632}

\bn\i
First example is a DP consisting of two SPs named $A$ and $B$, which is a model for transmitting one message $x$
from $A$ to $B$ through a hidden channel  $c_{AB}$ 
(only $A$ and $B$ know the name of this channel).  

This DP works as follows:
\bi\i
 $A$ sends  $B$ the message
$x$ through channel 
$c_{AB}$, 
\i  
$B$ receives a message from  channel $c_{AB}$,
writes this message to the variable $y$,  
and then it behaves in the same way as the SP $P$.
\ei

SPs $A$ and $B$ are defined as follows:
$$\by
A=(\bar c_{AB}! x).{\bf 0},
\quad
B=(\bar c_{AB}?\hat y).P.
\ey
$$

The scheme of the 
DP $(A,B)$ has the following form:
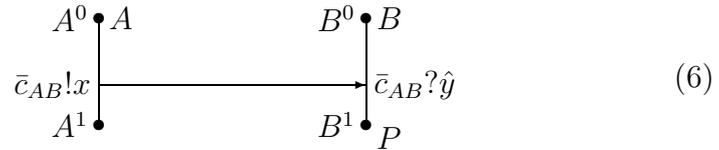
\be{primerdfadsljfk1}\by
\begin{picture}(100,40)
\put(0,15){\vector(1,0){100}}
\put(0,0){\line(0,1){40}}
\put(100,0){\line(0,1){40}}
\put(4,40){\makebox(0,0)[l]{$A$}}
\put(104,40){\makebox(0,0)[l]{$B$}}
\put(0,40){\circle*{4}}
\put(100,40){\circle*{4}}
\put(0,0){\circle*{4}}
\put(100,0){\circle*{4}}
\put(-4,40){\makebox(0,0)[r]{$A^0$}}
\put(-4,0){\makebox(0,0)[r]{$A^1$}}
\put(96,40){\makebox(0,0)[r]{$B^0$}}
\put(96,0){\makebox(0,0)[r]{$B^1$}}
\put(104,-5){\makebox(0,0)[l]{$P$}}
\put(-3,15){\makebox(0,0)[r]{$\bar c_{AB}!x$}}
\put(103,15){\makebox(0,0)[l]{$\bar c_{AB}?\hat y$}}
\end{picture}\ey
\ee

\i
Second example is a DP consisting of two SPs named $A$ and $B$, which is a model of transmission an encrypted message
$k_{AB}(x)$   from 
$A$ to $B$ 
through the open channel $\circ$. 
It is assumed that $A$ and $B$ have a shared secret 
key 
$k_{AB}$, on which they can encrypt and decrypt messages
using a symmetric encryption system, 
and only  $A$ and $B$ know the key  $k_{AB}$.

This DP works as follows:
\bi
\i $A$ sends $B$ an encrypted
 message $k_{AB}(x)$ to channel $\circ$,
\i $B$ receives the message $k_{AB}(x)$
from channel $\circ$,
decrypts it, writes  the extracted  message 
$x$ into  the variable $y$, and then behaves 
in the same way as SP $P$.
\ei

SPs $A$ and $B$ are defined as follows:
$$\by
A= (! \bar k_{AB}(x)).{\bf 0},
\quad
B=(?\bar k_{AB}(\hat y)).P.
\ey
$$

A scheme of DP $(A,B)$ has the following form:
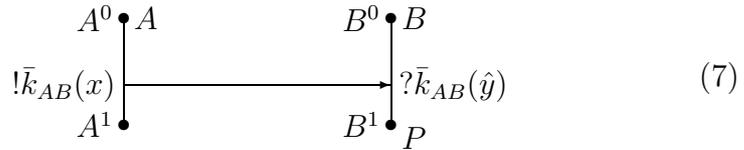
\be{primerdfadsljfk2}\by
\begin{picture}(100,40)
\put(0,40){\circle*{4}}
\put(100,40){\circle*{4}}
\put(0,0){\circle*{4}}
\put(100,0){\circle*{4}}
\put(-4,40){\makebox(0,0)[r]{$A^0$}}
\put(-4,0){\makebox(0,0)[r]{$A^1$}}
\put(96,40){\makebox(0,0)[r]{$B^0$}}
\put(96,0){\makebox(0,0)[r]{$B^1$}}
\put(0,15){\vector(1,0){100}}
\put(0,0){\line(0,1){40}}
\put(100,0){\line(0,1){40}}
\put(4,40){\makebox(0,0)[l]{$A$}}
\put(104,40){\makebox(0,0)[l]{$B$}}
\put(104,-5){\makebox(0,0)[l]{$P$}}
\put(-3,15){\makebox(0,0)[r]{$!\bar k_{AB}(x)$}}
\put(103,15){\makebox(0,0)[l]{$?\bar k_{AB}(\hat y)$}}
\end{picture}\ey
\ee

\i
Third example is a DP consisting of three SPs named $A$, $B$, and $T$, which is a model for transmission one message $x$ from $A$ to $B$  through a hidden channel $c_{AB}$,
 using a {\bf trusted intermediary} $T$, where
$A$ and $T$ ($B$ and $T$) communicate
 through a hidden channel $c_{AT}$ ($c_{BT}$), and only $A$ and $T$ ($B$ and $T$) know the name of this channel.

This DP works as follows:
\bi
\i
 $A$ sends $T$ channel name $c_{AB}$ 
 (only $A$ knows name $c_{AB}$ at first) 
 through channel $c_{AT}$, 
\i $T$  sends $B$  received channel name $c_{AB}$
through channel  $c_{BT}$,
\i
 $A$ sends $B$ message
$x$ through  channel 
$c_{AB}$, 
\i  
$B$ receives a message from  channel $c_{AB}$
and
writes it to variable  $y$  and then 
it behaves in the same way as SP $P$.
\ei

SPs $A$, $B$ and $T$ are defined as follows:
\be{sdfgdsfgdsfgw3rt4ge46h5urjhyr}
\by
A=  \alpha_1.
 \alpha_2.
{\bf 0},&\mbox{where}&  \alpha_1=\bar c_{AT}!\bar c_{AB},&
 \alpha_2=\bar c_{AB}!  x,\\
T=  \gamma_1.
 \gamma_2.{\bf 0},&\mbox{where}& 
 \gamma_1=
 \bar c_{AT}?\hat u,&
 \gamma_2=\bar c_{BT}! u,\\
B=  \beta_1.
 \beta_2.P,&\mbox{where}& 
 \beta_1= \bar c_{BT}?\hat v,&
 \beta_2=v?\hat y.
\ey
\ee

A scheme of DP $(A,B,T)$ has the following form:
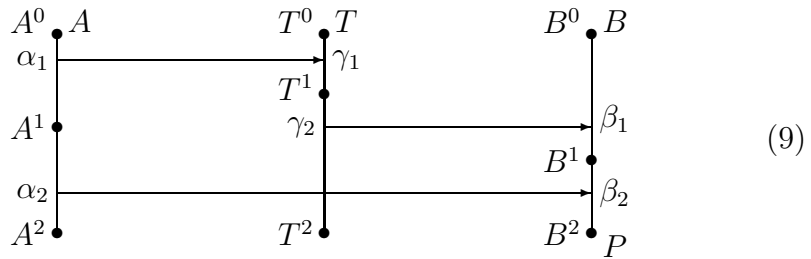
\be{primerdfadsljfk3}\by
\begin{picture}(0,85)
\put(-100,80){\circle*{4}}
\put(0,80){\circle*{4}}
\put(100,80){\circle*{4}}

\put(-100,45){\circle*{4}}
\put(0,57.5){\circle*{4}}
\put(100,32.5){\circle*{4}}

\put(-100,5){\circle*{4}}
\put(0,5){\circle*{4}}
\put(100,5){\circle*{4}}

\put(-104,85){\makebox(0,0)[r]{$A^0$}}
\put(-104,45){\makebox(0,0)[r]{$A^1$}}
\put(-104,5){\makebox(0,0)[r]{$A^2$}}

\put(-4,85){\makebox(0,0)[r]{$T^0$}}
\put(-4,60){\makebox(0,0)[r]{$T^1$}}
\put(-4,5){\makebox(0,0)[r]{$T^2$}}

\put(96,85){\makebox(0,0)[r]{$B^0$}}
\put(96,32.5){\makebox(0,0)[r]{$B^1$}}
\put(96,5){\makebox(0,0)[r]{$B^2$}}

\put(-100,70){\vector(1,0){100}}
\put(0,45){\vector(1,0){100}}
\put(-100,20){\vector(1,0){200}}
\put(100,5){\line(0,1){75}}
\put(0,5){\line(0,1){75}}
\put(-100,5){\line(0,1){75}}

\put(-96,85){\makebox(0,0)[l]{$A$}}
\put(4,85){\makebox(0,0)[l]{$T$}}
\put(104,85){\makebox(0,0)[l]{$B$}}
\put(104,0){\makebox(0,0)[l]{$P$}}
\put(-103,70){\makebox(0,0)[r]{$ \alpha_1$}}
\put(-3,45){\makebox(0,0)[r]{$ \gamma_2$}}
\put(3,70){\makebox(0,0)[l]{$ \gamma_1$}}
\put(-103,20){\makebox(0,0)[r]{$ \alpha_2$}}
\put(103,20){\makebox(0,0)[l]{$ \beta_2$}}
\put(103,48){\makebox(0,0)[l]{$ \beta_1$}}

\end{picture}\ey
\ee

\i
Fourth example is a DP (called a {\bf Wide-Mouth Frog (WMF)} protocol),  consisting of three SPs named $A$, $B$ and $T$
(where $T$ is a trusted intermediary).
This DP is a model of a transmission of encrypted message $k_{AB}(x)$ from $A$ to $B$ through open channel $\circ$  with use of $T$, with whom  $A$ and $B$ communicate through open channel $\circ$. SP $A$ \bi\i creates the secret key $k_{AB}$, \i sends $B$ this key in an encrypted form using $T$, and then \i sends $B$ encrypted message $k_{AB}(x)$.\ei

It is assumed that $A$ and $T$ ($B$ and $T$) have a shared secret key $k_{AT}$ ($k_{BT}$),  on which they can encrypt and decrypt messages
using a symmetric encryption system, 
and only 
$A$ and $T$ ($B$ and $T$) know secret key  $k_{AT}$
($k_{BT}$).

This DP works as follows.

\bi
\i  $A$ creates a secret key $k_{AB}$
(at first only $A$ knows this key)
and sends $T$ encrypted message $k_{AT}(k_{AB})$
through $\circ$,
then $A$ sends $B$ encrypted message $k_{AB}(x)$
through $\circ$, 
\i $T$ receives  a message from $A$, decrypts this message, then encrypts the extracted key $k_{AB}$ with the key $k_{BT}$, and sends $B$ encrypted message
$k_{BT}(k_{AB})$ through $\circ$,
\i  $B$ extracts  key $k_{AB}$ 
from the message received  from $T$, 
and then uses this key 
to extract  message $x$ from the message 
received from $A$, writes $x$ to  variable $y$, 
and then behaves in the same way as  SP
$P$.
\ei

SPs $A$, $B$ and $T$ are defined as follows:
\be{sdfgdsfgdsfgw3rt4ge46h5urjhyr1}
\by
A= \alpha_1.
\alpha_2.
{\bf 0},&\mbox{where}& \alpha_1=!\bar k_{AT}(\bar k_{AB}),&
\alpha_2=!\bar k_{AB}(x),\\
T= \gamma_1.
\gamma_2.{\bf 0},&\mbox{where}& \gamma_1=? \bar k_{AT}(\hat u),&
\gamma_2=!{\bar k_{BT}(u)},\\
B= \beta_1.
\beta_2.P,&\mbox{where}& \beta_1=? \bar k_{BT}(\hat v),&
\beta_2=? v(\hat y).
\ey
\ee

A scheme of  DP $(A,B,T)$ has the same  form
\re{primerdfadsljfk3}, as the scheme of the previous DP.
\en

\subsection{Transition graphs of  distributed processes}

\subsubsection{A concept of a transition graph
of a distributed process}
\label{dsfasfdsagfg}

Let $P$ be a DP of the form $\prod_{i\in I}P_i$ .

A {\bf transition graph (TG)} of DP $P$ 
is a graph $G_P$ such that
\bi
\i a set of nodes of $G_P$ is the Cartesian product 
of the sets of nodes of graphs from $P_I$, 
i.e. 
each node of  $G_P$ is a family of nodes 
$$V=\{v_i\mid i\in I\}, \mbox{ where }\forall\,i\in I\;\;v_i\in P_i,$$ \i each edge of $G_P$  has the form
\be{fdsgdsfghhdg43}
\{v_i\mid i\in I\}\ral{\!\alpha_{P_i}}\{v'_i\mid i\in I\},\ee
where
$P_i$ has the edge $v_i\ra{\alpha} v'_i$ and
$\forall\,i'\in I\setminus\{i\}\quad v_{i'}=v'_{i'}$.
\ei

The node
 $\{P^0_i\mid i\in I\}\in G_P$ is said to be
an {\bf initial} node of $G_P$, and is denoted by $G^0_P$.
An edge $V\ral{\!\alpha_{P_i}}V'$ is said to be a
{\bf realizable} edge, if $\exists\,S,S'\in \Sigma_P$:
$V=V^S$ and $V'=V^{S'}$.

It is not difficult to prove that if $\forall\,i\in I\;
P_i$ is acyclic, then $G_P$ is acyclic.

For each DP $P$ the graph $G_{P^*}$ can be considered 
as a completion of the graph $G_P$ with cyclic edges corresponding to the actions of $P_*$.

If DP  $P$ has the form  $(P_1,\ldots, P_n)$, 
then the following conventions will be used 
in a graphical representation of $G_P$:
\bi
\i each node
$V=\{v_i\mid i=1,\ldots, n\}$ of $G_P$
is represented by an oval, 
there is a list $v_1\ldots v_n$ 
of components of $V$ inside this oval, 
\i an initial node $G^0_P$ is represented by a double oval.
\ei

\subsubsection{Examples of  transition graphs 
of distributed processes}
\label{dfgdshghdfghdsfsd4}

In this section we outline some examples of TGs
for DPs described by  schemes from section
\ref{fadsgadsg3w54ggfd}.

\bn\i A TG for a DP described by  scheme
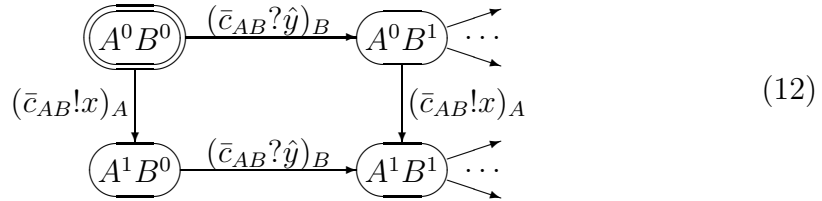
\re{primerdfadsljfk1}:
\be{zfdsas3453y46354y325221}
\by
\begin{picture}(150,65)
\put(0,50){\oval(34,20)}
\put(0,50){\makebox(0,0)[c]{${ A^0B^0}$}}
\put(100,50){\oval(34,20)}
\put(100,50){\makebox(0,0)[c]{${ A^0B^1}$}}
\put(0,0){\oval(34,20)}
\put(0,50){\oval(38,24)}
\put(0,0){\makebox(0,0)[c]{${A^1B^0}$}}
\put(100,0){\oval(34,20)}
\put(100,0){\makebox(0,0)[c]{${A^1B^1}$}}

\put(0,38){\vector(0,-1){28}}
\put(100,40){\vector(0,-1){30}}

\put(17,0){\vector(1,0){66}}
\put(19,50){\vector(1,0){64}}

\put(2,25){\makebox(0,0)[r]{
$ (\bar c_{AB}! x)_A$
}}
\put(98,25){\makebox(0,0)[l]{
$ (\bar c_{AB}! x)_A$
}}

\put(50,0){\makebox(0,0)[b]{
$ (\bar c_{AB}? \hat y)_B$
}}

\put(50,50){\makebox(0,0)[b]{
$ (\bar c_{AB}? \hat y)_B$
}}

\put(117,54){\vector(3,1){20}}
\put(117,46){\vector(3,-1){20}}
\put(130,50){\makebox(0,0){
$\ldots$
}}

\put(117,4){\vector(3,1){20}}
\put(117,-4){\vector(3,-1){20}}
\put(130,0){\makebox(0,0){
$\ldots$
}}
\end{picture}
\ey
\ee
\vspace{2mm}
where the slanted arrows denote  \bi\i edges of $G_P$ 
outgoing from the corresponding  nodes, \i and  parts of $G_P$ reachable  after passing through these edges, \ei which are not represented in this picture, 
this convention will be used in the following TG examples as well.

\i A TG for a DP described by scheme
\re{primerdfadsljfk2}:

\be{zfdsas3453y46354y3252211327}
\by
\begin{picture}(150,60)
\put(0,50){\oval(34,20)}
\put(0,50){\makebox(0,0)[c]{${ A^0B^0}$}}
\put(100,50){\oval(34,20)}
\put(100,50){\makebox(0,0)[c]{${ A^0B^1}$}}

\put(0,0){\oval(34,20)}
\put(0,50){\oval(38,24)}
\put(0,0){\makebox(0,0)[c]{${A^1B^0}$}}
\put(100,0){\oval(34,20)}
\put(100,0){\makebox(0,0)[c]{${A^1B^1}$}}

\put(0,38){\vector(0,-1){28}}
\put(100,40){\vector(0,-1){30}}

\put(17,0){\vector(1,0){66}}
\put(19,50){\vector(1,0){64}}

\put(2,25){\makebox(0,0)[r]{
$!\bar k_{AB}(x)$
}}
\put(98,25){\makebox(0,0)[l]{
$!\bar k_{AB}(x)$
}}

\put(50,0){\makebox(0,0)[b]{
$?\bar k_{AB}(\hat y)$
}}

\put(50,50){\makebox(0,0)[b]{
$?\bar k_{AB}(\hat y)$
}}

\put(117,54){\vector(3,1){20}}
\put(117,46){\vector(3,-1){20}}
\put(130,50){\makebox(0,0){
$\ldots$
}}
\put(117,4){\vector(3,1){20}}
\put(117,-4){\vector(3,-1){20}}
\put(130,0){\makebox(0,0){
$\ldots$}}

\end{picture}
\ey
\ee
\vspace{2mm}

 \i A TG for a DP described by scheme
\re{primerdfadsljfk3}:

{\def\arraystretch{1}
{\small
\be{zfdsas3453y46354y51}
\by
\begin{picture}(150,270)

\put(117,4){\vector(3,1){20}}
\put(117,-4){\vector(3,-1){20}}
\put(130,0){\makebox(0,0){$\ldots$}}
\put(147,34){\vector(3,1){20}}
\put(147,26){\vector(3,-1){20}}
\put(160,30){\makebox(0,0){$\ldots$}}
\put(177,64){\vector(3,1){20}}
\put(177,56){\vector(3,-1){20}}
\put(190,60){\makebox(0,0){$\ldots$}}

\put(117,104){\vector(3,1){20}}
\put(117,96){\vector(3,-1){20}}
\put(130,100){\makebox(0,0){$\ldots$}}
\put(147,134){\vector(3,1){20}}
\put(147,126){\vector(3,-1){20}}
\put(160,130){\makebox(0,0){$\ldots$}}
\put(177,164){\vector(3,1){20}}
\put(177,156){\vector(3,-1){20}}
\put(190,160){\makebox(0,0){$\ldots$}}

\put(117,204){\vector(3,1){20}}
\put(117,196){\vector(3,-1){20}}
\put(130,200){\makebox(0,0){$\ldots$}}
\put(147,234){\vector(3,1){20}}
\put(147,226){\vector(3,-1){20}}
\put(160,230){\makebox(0,0){$\ldots$}}
\put(177,264){\vector(3,1){20}}
\put(177,256){\vector(3,-1){20}}
\put(190,260){\makebox(0,0){$\ldots$}}

\put(-100,200){\oval(34,20)}
\put(-100,200){\oval(38,24)}
\put(-100,200){\makebox(0,0)[c]{${\scriptstyle A^0T^0B^0}$}}
\put(0,200){\oval(34,20)}
\put(0,200){\makebox(0,0)[c]{${\scriptstyle A^0T^0B^1}$}}
\put(100,200){\oval(34,20)}
\put(100,200){\makebox(0,0)[c]{${\scriptstyle A^0T^0B^2}$}}

\put(-70,230){\oval(34,20)}
\put(-70,230){\makebox(0,0)[c]{${\scriptstyle A^0T^1B^0}$}}
\put(30,230){\oval(34,20)}
\put(30,230){\makebox(0,0)[c]{${\scriptstyle A^0T^1B^1}$}}
\put(130,230){\oval(34,20)}
\put(130,230){\makebox(0,0)[c]{${\scriptstyle A^0T^1B^2}$}}

\put(-40,260){\oval(34,20)}
\put(-40,260){\makebox(0,0)[c]{${\scriptstyle A^0T^2B^0}$}}
\put(60,260){\oval(34,20)}
\put(60,260){\makebox(0,0)[c]{${\scriptstyle A^0T^2B^1}$}}
\put(160,260){\oval(34,20)}
\put(160,260){\makebox(0,0)[c]{${\scriptstyle A^0T^2B^2}$}}

\put(-100,100){\oval(34,20)}
\put(-100,100){\makebox(0,0)[c]{${\scriptstyle A^1T^0B^0}$}}
\put(0,100){\oval(34,20)}
\put(0,100){\makebox(0,0)[c]{${\scriptstyle A^1T^0B^1}$}}
\put(100,100){\oval(34,20)}
\put(100,100){\makebox(0,0)[c]{${\scriptstyle A^1T^0B^2}$}}

\put(-70,130){\oval(34,20)}
\put(-70,130){\makebox(0,0)[c]{${\scriptstyle A^1T^1B^0}$}}
\put(30,130){\oval(34,20)}
\put(30,130){\makebox(0,0)[c]{${\scriptstyle A^1T^1B^1}$}}
\put(130,130){\oval(34,20)}
\put(130,130){\makebox(0,0)[c]{${\scriptstyle A^1T^1B^2}$}}

\put(-40,160){\oval(34,20)}
\put(-40,160){\makebox(0,0)[c]{${\scriptstyle A^1T^2B^0}$}}
\put(60,160){\oval(34,20)}
\put(60,160){\makebox(0,0)[c]{${\scriptstyle A^1T^2B^1}$}}
\put(160,160){\oval(34,20)}
\put(160,160){\makebox(0,0)[c]{${\scriptstyle A^1T^2B^2}$}}

\put(-100,0){\oval(34,20)}
\put(-100,0){\makebox(0,0)[c]{${\scriptstyle A^2T^0B^0}$}}
\put(0,0){\oval(34,20)}
\put(0,0){\makebox(0,0)[c]{${\scriptstyle A^2T^0B^1}$}}
\put(100,0){\oval(34,20)}
\put(100,0){\makebox(0,0)[c]{${\scriptstyle A^2T^0B^2}$}}

\put(-70,30){\oval(34,20)}
\put(-70,30){\makebox(0,0)[c]{${\scriptstyle A^2T^1B^0}$}}
\put(30,30){\oval(34,20)}
\put(30,30){\makebox(0,0)[c]{${\scriptstyle A^2T^1B^1}$}}
\put(130,30){\oval(34,20)}
\put(130,30){\makebox(0,0)[c]{${\scriptstyle A^2T^1B^2}$}}

\put(-40,60){\oval(34,20)}
\put(-40,60){\makebox(0,0)[c]{${\scriptstyle A^2T^2B^0}$}}
\put(60,60){\oval(34,20)}
\put(60,60){\makebox(0,0)[c]{${\scriptstyle A^2T^2B^1}$}}
\put(160,60){\oval(34,20)}
\put(160,60){\makebox(0,0)[c]{${\scriptstyle A^2T^2B^2}$}}

\put(-100,188){\vector(0,-1){78}}
\put(0,190){\vector(0,-1){80}}
\put(100,190){\vector(0,-1){80}}
\put(-100,90){\vector(0,-1){80}}
\put(0,90){\vector(0,-1){80}}
\put(100,90){\vector(0,-1){80}}

\put(-70,220){\vector(0,-1){80}}
\put(30,220){\vector(0,-1){80}}
\put(130,220){\vector(0,-1){80}}
\put(-70,120){\vector(0,-1){80}}
\put(30,120){\vector(0,-1){80}}
\put(130,120){\vector(0,-1){80}}

\put(-40,250){\vector(0,-1){80}}
\put(60,250){\vector(0,-1){80}}
\put(160,250){\vector(0,-1){80}}
\put(-40,150){\vector(0,-1){80}}
\put(60,150){\vector(0,-1){80}}
\put(160,150){\vector(0,-1){80}}

\put(-81,200){\vector(1,0){64}}
\put(17,200){\vector(1,0){66}}
\put(-83,100){\vector(1,0){66}}
\put(17,100){\vector(1,0){66}}
\put(-83,0){\vector(1,0){66}}
\put(17,0){\vector(1,0){66}}

\put(-53,230){\vector(1,0){66}}
\put(47,230){\vector(1,0){66}}
\put(-53,130){\vector(1,0){66}}
\put(47,130){\vector(1,0){66}}
\put(-53,30){\vector(1,0){66}}
\put(47,30){\vector(1,0){66}}

\put(-23,260){\vector(1,0){66}}
\put(77,260){\vector(1,0){66}}
\put(-23,160){\vector(1,0){66}}
\put(77,160){\vector(1,0){66}}
\put(-23,60){\vector(1,0){66}}
\put(77,60){\vector(1,0){66}}

\put(-93,212){\vector(1,1){10}}
\put(10,210){\vector(1,1){10}}
\put(110,210){\vector(1,1){10}}
\put(-60,240){\vector(1,1){10}}
\put(40,240){\vector(1,1){10}}
\put(140,240){\vector(1,1){10}}

\put(-90,110){\vector(1,1){10}}
\put(10,110){\vector(1,1){10}}
\put(110,110){\vector(1,1){10}}
\put(-60,140){\vector(1,1){10}}
\put(40,140){\vector(1,1){10}}
\put(140,140){\vector(1,1){10}}

\put(-90,10){\vector(1,1){10}}
\put(10,10){\vector(1,1){10}}
\put(110,10){\vector(1,1){10}}
\put(-60,40){\vector(1,1){10}}
\put(40,40){\vector(1,1){10}}
\put(140,40){\vector(1,1){10}}

\put(-97,150){\makebox(0,0)[r]{
$\alpha_1$
}}
\put(3,150){\makebox(0,0)[r]{
$\alpha_1$
}}
\put(103,150){\makebox(0,0)[r]{
$\alpha_1$
}}

\put(-67,180){\makebox(0,0)[r]{
$\alpha_1$
}}
\put(33,180){\makebox(0,0)[r]{
$\alpha_1$
}}
\put(133,180){\makebox(0,0)[r]{
$\alpha_1$
}}

\put(-42,210){\makebox(0,0)[l]{
$\alpha_1$
}}
\put(58,210){\makebox(0,0)[l]{
$\alpha_1$
}}
\put(158,210){\makebox(0,0)[l]{
$\alpha_1$
}}

\put(-97,50){\makebox(0,0)[r]{
$\alpha_2$
}}
\put(3,50){\makebox(0,0)[r]{
$\alpha_2$
}}
\put(103,50){\makebox(0,0)[r]{
$\alpha_2$
}}

\put(-67,80){\makebox(0,0)[r]{
$\alpha_2$
}}
\put(33,80){\makebox(0,0)[r]{
$\alpha_2$
}}
\put(133,80){\makebox(0,0)[r]{
$\alpha_2$
}}

\put(-42,110){\makebox(0,0)[l]{
$\alpha_2$
}}
\put(58,110){\makebox(0,0)[l]{
$\alpha_2$
}}
\put(158,110){\makebox(0,0)[l]{
$\alpha_2$
}}

\put(-50,0){\makebox(0,0)[b]{
$\beta_1$
}}
\put(50,0){\makebox(0,0)[b]{
$\beta_2$
}}
\put(-20,30){\makebox(0,0)[b]{
$\beta_1$
}}
\put(70,30){\makebox(0,0)[b]{
$\beta_2$
}}
\put(10,60){\makebox(0,0)[b]{
$\beta_1$
}}
\put(110,60){\makebox(0,0)[b]{
$\beta_2$
}}

\put(-50,100){\makebox(0,0)[b]{
$\beta_1$
}}
\put(50,100){\makebox(0,0)[b]{
$\beta_2$
}}
\put(-20,130){\makebox(0,0)[b]{
$\beta_1$
}}
\put(70,130){\makebox(0,0)[b]{
$\beta_2$
}}
\put(10,160){\makebox(0,0)[b]{
$\beta_1$
}}
\put(110,160){\makebox(0,0)[b]{
$\beta_2$
}}

\put(-50,200){\makebox(0,0)[b]{
$\beta_1$
}}
\put(50,200){\makebox(0,0)[b]{
$\beta_2$
}}
\put(-20,230){\makebox(0,0)[b]{
$\beta_1$
}}
\put(70,230){\makebox(0,0)[b]{
$\beta_2$
}}
\put(10,260){\makebox(0,0)[b]{
$\beta_1$
}}
\put(110,260){\makebox(0,0)[b]{
$\beta_2$
}}

\put(-82,17){\makebox(0,0)[r]{
$\gamma_1$
}}
\put(18,17){\makebox(0,0)[r]{
$\gamma_1$
}}
\put(118,17){\makebox(0,0)[r]{
$\gamma_1$
}}

\put(-82,117){\makebox(0,0)[r]{
$\gamma_1$
}}
\put(18,117){\makebox(0,0)[r]{
$\gamma_1$
}}
\put(118,117){\makebox(0,0)[r]{
$\gamma_1$
}}

\put(-84,219){\makebox(0,0)[r]{
$\gamma_1$
}}
\put(18,217){\makebox(0,0)[r]{
$\gamma_1$
}}
\put(118,217){\makebox(0,0)[r]{
$\gamma_1$
}}

\put(-52,47){\makebox(0,0)[r]{
$\gamma_2$
}}
\put(48,47){\makebox(0,0)[r]{
$\gamma_2$
}}
\put(148,47){\makebox(0,0)[r]{
$\gamma_2$
}}

\put(-52,147){\makebox(0,0)[r]{
$\gamma_2$
}}
\put(48,147){\makebox(0,0)[r]{
$\gamma_2$
}}
\put(148,147){\makebox(0,0)[r]{
$\gamma_2$
}}

\put(-52,247){\makebox(0,0)[r]{
$\gamma_2$
}}
\put(48,247){\makebox(0,0)[r]{
$\gamma_2$
}}
\put(148,247){\makebox(0,0)[r]{
$\gamma_2$
}}

\end{picture}
\ey
\ee
 }
}\vspace{5mm}

\en

 \subsection{Values of expressions and formulas in states of  distributed processes}

 \subsubsection{A concept of a value of an expression and a formula in a state of a distributed process}

Let there are given the DP
$P=\prod_{i\in I}P_i$,
the state $S\in\Sigma_P$,
the expression $E\in Expr$, 
and the formula $\beta\in Fm$.

The notation $E^S$ denotes a subset of the set
 $Tm$, called a {\bf value of the expression
$E$ in the state $S$}, and defined as follows:
\bi
\i if $E\subseteq Tm$, then $E^S = E^{\theta^S}$,
\i if $E=X_P$, then $E^S=(X_P^S)^{\theta^S}$,
\i if $E=M_c$, then $E^S=M_{c^{\theta^S}}^S$,
\i if $E=k^{-1}(E')$, then $E^S=\{e\in Tm\mid
\exists\,e'\in (E')^S:
k^{\theta^S}(e)\subseteq e'
\}$,

\i
$(E\cap E')^S=E^S\cap (E')^S$,
$(E\cup E')^S=E^S\cup (E')^S$,
$(\neg E)^S=Tm\setminus E^S$.
\ei

The notation 
$S\models \beta$ denotes the statement 
{\bf $\beta$ holds in $S$}, which
is true iff one of the following cases holds:
\bi\i \bi\i
$\beta =(E=E')$, $(E\subseteq E')$, or $(E\supseteq E')$,
where $E,E'\in Expr$, and
\i $E^S=(E')^{S}$, 
$E^S\subseteq (E')^{S}$, or
$E^S\supseteq (E')^{S}$,  respectively,
\ei
\i \bi\i $\beta=(x\,\bot\,P_i)$, where 
$x\in
 {\it Var}$, $i\in I$, and \i
 $\forall\,e\in (X_{P_i}^S)^{\theta^S}\;\; 
 x\not\in {\it Var}_e,$\ei
  \i \bi\i
 $\beta=(x\,\bot\,C)$, 
where $x\in
 {\it Var}$, $C\subseteq Channels$, and\i
 $\forall\,c\in C,\;\forall\,e\in M^S_c\;\;
x\not\in {\it Var}_e,$\ei
\i \bi\i $\beta=(k\,\bot_{\bf K}\,P_i)$, where 
$k\in
 {\it Keys}$, $i\in I$, and \i
 $\forall\,e\in (X_{P_i}^S)^{\theta^S}$
each occurrence  of $k$ in $e$ is hidden,\ei
  \i \bi\i
 $\beta=(k\,\bot_{\bf K}\,C)$, 
where $k\in
 {\it Keys}$, $C\subseteq Channels$, and\i
 $\forall\,c\in C,\;\forall\,e\in M^S_c$
each occurrence  of  $k$ in $e$ is hidden,
\ei
\i $\beta=\{\beta_i\mid i\in I\}$ is a family of EFs,
$\forall\,i\in I\;\;S\models\beta_i$. 
\ei

\subsubsection{Theorems on preserving 
 values of formulas under transitions}
\label{sadfaaerg}

Below we prove  theorems that some formulas have the same values in states related by a transition relation.\\

\refstepcounter{theorem}
{\bf Theorem \arabic{theorem}\label{doredtuslnab322eq}}.

Let $P=\prod_{i\in I}P_i$ be a DP 
and  $S,S'\in \Sigma_P$ be states such that
$$\exists\,i\in I:\;
S\ral{\!\alpha_{P_i}}S'.$$

Then the implication 
$S\models \beta\;\Rightarrow\;S'\models \beta$
holds,
where $\beta$ is a formula of one of the following forms:
\bn
\i
$\beta=\{x\,\bot\,P_i, 
x\,\bot\,Channels
\}$, where $x\in X_{P}$,
\i
$\beta=\{k\,\bot_{\bf K}\,P_i, 
k\,\bot_{\bf K}\,Channels
\}$, where $k\in  X_{P_i}^{\bf K}$.
\en

{\bf Proof}.

\bn\i Let $\beta=\{x\,\bot\,P_i, 
x\,\bot\,Channels
\}$, where $x\in  X_{P}$.

 $S\models \beta$ means  that
\be{fdghsdfsg434}\left.
\by
\forall\,y\in X_{P_i}^S\quad
x\not\in {\it Var}_{y^{\theta^{S}}}\\
\forall\,c\in Channels,\forall\,e\in M^S_c\quad
x\not\in {\it Var}_e.
\ey\right\}\ee

It is required to prove that 
\re{fdghsdfsg434} implies $S'\models \beta$, i.e.
\be{fdghsdfsg43234}\left.
\by
\forall\,y\in X_{P_i}^{S'}\quad
x\not\in {\it Var}_{y^{\theta^{S'}}}\\
\forall\,c\in Channels,\forall\,e\in M^{S'}_c\quad
x\not\in {\it Var}_e.
\ey\right\}\ee

If first statement in \re{fdghsdfsg43234} is wrong, 
then  first statement in \re{fdghsdfsg434} 
 implies that $X_{P_i}^S\neq X_{P_i}^{S'}$. 
This is only possible if
\be{asfdgagdsfgfsg34}
\by 
\mbox{
$\alpha$ is of the form  $c?e$, 
 $X_{P_i}^{S'} = X_{P_i}^S\cup 
 {\it Var}_e$,}\\
\mbox{ $e^{\theta^{S'}}\in M^S_{c^{\theta^S}}$,
and $\exists\,y\in {\it Var}_e: x\in y^{\theta^{S'}}\;\;
(\Rightarrow\; x \in {\it Var}_{e^{\theta^{S'}}})$.}
\ey
\ee

\re{asfdgagdsfgfsg34} contradicts 
second statement in  \re{fdghsdfsg434}.

If  second statement in \re{fdghsdfsg43234}
is wrong, then second statement in
\re{fdghsdfsg434} implies that $\exists\,c\in Channels: M^S_c\neq M^{S'}_{c}$.
This is only possible if \be{sdfdasg45y4u6y5}
\left.\by
\mbox{$\alpha$
has the form $c'!e$, where $(c')^{\theta^S}=c$,
and $e\in Tm(X_{P_i}^S)$,}\\
\mbox{$M^{S'}_{c} = M^S_c\cup \{e^{\theta^{S}}\}$, 
and $x\in {\it Var}_{e^{\theta^{S}}}$.}
\ey\right\}\ee

Denote by symbols  $X$ and $\theta$ 
the set $X_{P_i}^S$ and the binding $\theta^{S}$,
respectively. 
From \re{sdfdasg45y4u6y5} it follows that
$e\in Tm(X)$ and 
$x\in {\it Var}_{e^\theta}$.

From $x\in {\it Var}_{e^\theta}$ it follows that
$\exists\,y\in {\it Var}_e: x\in {\it Var}_{y^\theta}$.

From $e\in Tm(X)$ and $y\in {\it Var}_e$
it follows that 
$y\in X$, so
$y^\theta\in X^\theta$. 

Thus, we get the statements
$$y^\theta\in X^\theta,\;\;x\in {\it Var}_{y^\theta}$$
that contradict first statement in
\re{fdghsdfsg434}.

\i Let $\beta=\{k\,\bot_{\bf K}\,P_i, 
k\,\bot_{\bf K}\,Channels
\}$, where $k\in  X_{P_i}^{\bf K}$.

$S\models \beta$ means  that
\be{fdghsdfsg4341}\!\!\!\!\!\!\!\!\!\!\!\!\!
\left.\by
\mbox{ $\forall\,x\in X_{P_i}^S$
each occurrence of $k$ in $x^{\theta^S}$ is
hidden,}
\\
\mbox{ $\forall\,c\in Channels,\;\forall\,e\in M^S_c$
each occurrence of $k$ in $e$ is hidden.}\\
\ey\right\}\ee

It is required to prove that  \re{fdghsdfsg4341}
implies $S'\models \beta$, i.e.
\be{fdghsdfsg432341}\!\!\!\!\!\!\!\!\!\!\!\!\!
\left.
\by
\mbox{ $\forall\,x\in X_{P_i}^{S'}$
each occurrence  of $k$ in $x^{\theta^{S'}}$ is 
hidden,}\\
\mbox{ $\forall\,c\in Channels,\;\forall\,e\in M^{S'}_c$ 
each occurrence  of $k$ in $e$ is hidden.}\\
\ey\right\}\ee

If  first statement in \re{fdghsdfsg432341}
is wrong, then  first statement in 
\re{fdghsdfsg4341} implies that
$X_{P_i}^S\neq
X_{P_i}^{S'}$.
This is possible in the following two cases:
\bn\i\label{adsgfdsghsdfgdsa}
$\left\{\by 
\mbox{
$\alpha$ is of the form $c?e$, 
 $X_{P_i}^{S'} = X_{P_i}^S\cup {\it Var}_e$,}\\
\mbox{ $e^{\theta^{S'}}\in M^S_{c^{\theta^S}}$, and 
$\exists\,y\in {\it Var}_e$:}\\\hspace{20mm}\mbox{
$\exists$ unhidden occurrence of
$\mbox{$k$ in $y^{\theta^{S'}}$,}$}
\ey\right.
$
\i\label{adsgfdsghsdfgdsa32}
$\left\{\by 
\mbox{
$\alpha$ is of the form $e:=e'$, 
 $X_{P_i}^{S'} = X_{P_i}^S\cup {\it Var}_e$,}\\
\mbox{ $e^{\theta^{S'}}=(e')^{\theta^S}$, and
$\exists\,y\in {\it Var}_e$:
}\\\hspace{20mm}\mbox{
$\exists$ unhidden occurrence of
$\mbox{$k$ in $y^{\theta^{S'}}$.}$}
\ey\right.
$
\en

In case \ref{adsgfdsghsdfgdsa} 
$\exists$ unhidden occurrence of $k$ in  
$e^{\theta^{S'}}$,
that contradicts second statement 
in \re{fdghsdfsg4341}.

In case \ref{adsgfdsghsdfgdsa32}  the following is true:
\be{fgsadf243213}
\mbox{$\exists\,e'\in Tm(X_{P_i}^S):
\exists$ unhidden occurrence of 
 $k$ in $(e')^{\theta^S}$.}\ee

However, according to  first statement in \re{fdghsdfsg4341}, 
$\forall\,x\in X_{P_i}^S$ each occurrence of $k$ in $x^{\theta^S}$ is hidden, whence by induction 
on the structure of $e'$ it is easy to prove that 
\re{fgsadf243213} is false.

If  second statement in  \re{fdghsdfsg432341} is wrong, then  second statement in \re{fdghsdfsg4341} implies that $\exists\,c\in Channels: M^S_c\neq M^{S'}_{c}$.
This is only possible if \be{sdfdasg45y4u6y51}
\!\!\!\!\!\!\!\!\by
\mbox{$\alpha$  has the form
$c'!e$, where $(c')^{\theta^S}=c$,
and $e\in Tm(X_{P_i}^S)$,}\\ 
\mbox{$M^{S'}_{c} = M^S_c\cup \{e^{\theta^{S}}\}$, 
and $\exists$ an unhidden occurrence  of
$k$ in $e^{\theta^{S}}$.}
\ey\ee

As in  previous case, we prove 
by induction on the structure of  $e$ 
that each occurrence  of
$k$ in $e^{\theta^S}$ is hidden 
(for the base of induction we use first statement in
\re{fdghsdfsg4341}) that 
contradicts the last statement in 
\re{sdfdasg45y4u6y51}.
 $\blackbox$
\en

\refstepcounter{theorem}
{\bf Theorem \arabic{theorem}\label{doredtuslna32b322eq}}.

Let $P=\prod_{i\in I}P_i$ be a DP, 
and $S,S'\in \Sigma_P$ be states such that
$$\exists\,i\in I:\;
S\ral{\!\alpha_{P_i}}S'.$$

Then the implication $S\models \beta\;\Rightarrow\;S'\models \beta$ holds,
where $\beta$ is a formula of one of the following two forms:
\be{fdgdsgsdfdsh4t4}
\{c\,\bot\,P_i, 
c\,\bot\,Channels, M_c= E
\},\mbox{ where  $c\in X_{P}^{\bf C}$,   }
\ee
\be{fdgdsgsdfdsh4t41}\!\!\!\!\!\!\!\!
\left\{
\by 
k\,\bot_{\bf K}\,P_i ,\;
k\,\bot_{\bf K}\,Channels, \\
k^{-1}(M_c)\subseteq E\quad(\forall\,c\in Channels),\\
k^{-1}(X_{P_i})\subseteq E
\ey
\right\},\mbox{where }  k\in  X_{P}^{\bf K},
E
\subseteq Tm(X_{P}).
\ee

{\bf Proof}.
\bn
\i 
Let  
$\beta$ has the form 
\re{fdgdsgsdfdsh4t4}.

According to theorem \ref{doredtuslnab322eq}, 
if first two EFs occurred in $\beta$ hold in $S$, 
then these EFs hold in $S'$ as well.

Thus, to prove $S'\models \beta$ 
it suffices to prove the implication
\be{sfddsgdshgfdsghdsf}
S\models 
\{c\,\bot\,P_i, 
c\,\bot\,Channels,M_c=E\}
\;\;\Rightarrow\;\;
S'\models  M_c=E.\ee

If the conclusion of implication
\re{sfddsgdshgfdsghdsf} does not hold,
then the sets
$M^S_c$ and $M^{S'}_c$ are different. 
This is possible only if 
$\alpha$ is of the form $c'!e$, where $c=(c')^{\theta^S}$
and $c'\in X^S_{P_i}$.
However, $S\models c\,\bot\,P_i$ implies that
 $c\not\in {\it Var}_{(c')^{\theta^S}}$,
i. e. $c\not\in \{c\}$, which is impossible.

\i 
Let $\beta$ has the form \re{fdgdsgsdfdsh4t41}.

According to  theorem \ref{doredtuslnab322eq}, 
if first two EFs occurred in $\beta$ hold in $S$, 
then these EFs hold in $S'$ as well.

Thus, to prove $S'\models \beta$ 
it suffices to prove the implication
\be{s3fddsgdshgfdsg3hdsf}\by
\left.\by
S\models k^{-1}(M_c)\subseteq E\quad(\forall\,c\in Channels)\\S\models k^{-1}(X_{P_i})\subseteq E\\
S\models 
\{k\,\bot_{\bf K}\,P_i, 
k\,\bot_{\bf K}\,Channels\}\ey
\right\}\Rightarrow\\
\Rightarrow
\left\{\by
S'\models k^{-1}(M_c)\subseteq E\quad(\forall\,c\in Channels)\\
S'\models k^{-1}(X_{P_i})\subseteq E
\ey\right.\ey\ee

\bn
\i
If  first statement in the conclusion of  implication  \re{s3fddsgdshgfdsg3hdsf} is wrong, then $\exists\,c\in Channels:
S'\not\models k^{-1}(M_c) \subseteq E$. 

From first statement  in the premise of implication \re{s3fddsgdshgfdsg3hdsf} 
 it follows that this is  possible
only if $$\by
\mbox{$\alpha$
has the form  $c'!e'$, where 
 $c=(c')^{\theta^{S}},
 e'\in Tm(X_{P_i}^S)$, }\\
\mbox{$
M^{S'}_{c} = M^S_c\cup \{(e')^{\theta^{S}}\}$, 
with
$\exists \,k(e)\subseteq (e')^{\theta^{S}}$:
$e\not\in  E$.}
\ey$$

The term $e'$ does not contain $k$, 
because  $e'\in Tm(X_{P_i}^S)$,
and if $e'$ contains $k$, then  $k\in X_{P_i}^S$,
which contradicts the assumption 
 $S\models k\,\bot_{\bf K}\,P_i$ in the premise of 
  implication 
\re{s3fddsgdshgfdsg3hdsf}.

Thus, $\exists\,x\in {\it Var}_{e'}\subseteq 
X_{P_i}^S$,
$\exists \,k(e)\subseteq x^{\theta^{S}}$,
and $e\not\in E$.

However, this contradicts the statement
$S\models k^{-1}(X_{P_i})\subseteq E$
in the premise of  implication  \re{s3fddsgdshgfdsg3hdsf}.
\i
If second statement in the conclusion of  implication 
\re{s3fddsgdshgfdsg3hdsf}
does not hold, 
then from second statement in the premise of implication
\re{s3fddsgdshgfdsg3hdsf} it follows that
$X_{P_i}^S\neq
X_{P_i}^{S'}$, and
\be{adsfadsfsfgewrgfgwf3343}
\exists\,x\in X_{P_i}^{S'}:
 \exists\,e\not\in E: k(e)\subseteq x^{\theta^{S'}}.\ee
 
This is possible in two cases:
\bn
\i\label{adsgfdsghsdfgdsa}
$\alpha=c?e'$, in this case
\be{zdfbdghgfhgfhsdhdr}
X_{P_i}^{S'} = X_{P_i}^S\cup {\it Var}_{e'},\;\; 
x\in {\it Var}_{e'},\;\;  
(e')^{\theta^{S'}}\in M^S_{c^{\theta^S}}.\ee
 
 Let  $c'=c^{\theta^S}$.
 
According to  first statement in the premise of  implication 
\re{s3fddsgdshgfdsg3hdsf}, 
$S\models k^{-1}(M_{c'})\subseteq E$,
so the following implication holds:
 \be{zdfgdsfgsg45ye5w}k(e)\subseteq \tilde e \in M^S_{c'}\;\Rightarrow\;e\in E.\ee
  
 The premise of  implication  \re{zdfgdsfgsg45ye5w}
holds when $\tilde e=(e')^{\theta^{S'}}$,
this follows from the last statement in \re{zdfbdghgfhgfhsdhdr} and from 
$$k(e)\subseteq x^{\theta^{S'}},\;\;x\in {\it Var}_{e'},\;\;
x^{\theta^{S'}}\subseteq (e')^{\theta^{S'}}\in M^S_{c^{\theta^S}}.$$

Thus, the conclusion of  implication
\re{zdfgdsfgsg45ye5w}
holds, which contradicts the statement
$e\not\in E$ in \re{adsfadsfsfgewrgfgwf3343}.

\i\label{adsgfdsghsdfgdsa32}

$\alpha=(e':=e'')$, in this case
\be{zdfbdghgfhgfhsdhdr}\by
X_{P_i}^{S'} = X_{P_i}^S\cup {\it Var}_{e'},\;\; 
x\in {\it Var}_{e'},\\  e''\in Tm(X^S_{P_i}),\;\;
(e')^{\theta^{S'}}=(e'')^{\theta^S}.\ey\ee

According to  second statement 
in the premise of implication
\re{s3fddsgdshgfdsg3hdsf}, 
$S\models k^{-1}(X_{P_i})\subseteq E$,
so the following implication holds:
 \be{zdfg23dsfgsg45ye5w}
 k(e)\subseteq \tilde e \in (X_{P_i}^S)^{\theta^S}\;\Rightarrow\;e\in E.\ee 

Since $x\in {\it Var}_{e'}$, then $$x^{\theta^{S'}}\subseteq 
(e')^{\theta^{S'}}=(e'')^{\theta^{S}}.$$

The last statements and \re{adsfadsfsfgewrgfgwf3343} 
imply the statements
\be{s3fddsgdshgfdsg3hdsf1} 
k(e)\subseteq  
(e'')^{\theta^{S}}\in (Tm(X^S_{P_i}))^{\theta^S}.\ee

The term $e''$ does not contain $k$, 
because $e''\in Tm(X_{P_i}^S)$,
and if $e''$ contains $k$, then  $k\in X_{P_i}^S$,
which contradicts the assumption $S\models k\,\bot_{\bf K}\,P_i$ 
in the premise of implication
\re{s3fddsgdshgfdsg3hdsf}. 

Hence, based on 
\re{s3fddsgdshgfdsg3hdsf1}, we obtain  
\be{dsafsf3w3w5t5367ui8}
\exists\,y\in {\it Var}_{e''}\subseteq 
X_{P_i}^S: k(e)\subseteq y^{\theta^{S}}.\ee

From \re{dsafsf3w3w5t5367ui8}
 it follows that if we define 
$\tilde e$ as the term $y^{\theta^{S}}$,
then the premise of implication
\re{zdfg23dsfgsg45ye5w} will be true. 

Consequently, a conclusion of this implication
will also be true, i.e. the statement
$e\in E$ is true, which contradicts the assumption 
 $e\not \in E$ in \re{adsfadsfsfgewrgfgwf3343}.
 $\blackbox$
\en\en\en

\refstepcounter{theorem}
{\bf Theorem \arabic{theorem}\label{doredtuslna32b322e3q}}.

Formula \re{fdgdsgsdfdsh4t41} in theorem 
\ref{doredtuslna32b322eq}
can be replaced by a formula $\beta$ of the form
\be{fdgdsgsdfdsh4t412}
\left\{
\by 
k\,\bot_{\bf K}\,P_i ,\;
k\,\bot_{\bf K}\,Channels, \\
k^{-1}(M_{c_0})= E\\
k^{-1}(M_c)\subseteq E\quad(\forall\,c\in Channels),\\
k^{-1}(X_{P_i})\subseteq E
\ey
\right\}\ee
where $c_0\in X_P^{\bf C}$,
$k\in  X_{P}^{\bf K},
E
\subseteq Tm(X_{P})$.\\

{\bf Proof}.

If $S\models \beta$, then
 $S\models \beta'$, where $\beta'$ is obtained from  
  $\beta$ 
by removing the formula $k^{-1}(M_{c_0})=E$.

According to theorem \ref{doredtuslna32b322eq},
the statement $S\models \beta'$ implies the statement
$S'\models \beta'$. In particular,
$S'\models k^{-1}(M_{c_0})\subseteq  E$.

\re{sdfagr33gr356ujyhfgd} implies the inclusion
$M^S_{c_0}\subseteq M^{S'}_{c_0}$, 
from which we obtain  the statements
$$E=(k^{-1}(M_{c_0}))^S\subseteq 
(k^{-1}(M_{c_0}))^{S'}\subseteq E,$$
therefore, $S'\models k^{-1}(M_{c_0})=  E$.
Thus, $S'\models \beta$.
$\blackbox$

\subsection{Marking  of a transition graph}

\subsubsection{A concept of a marking of a transition graph}

Let $P$ be a DP of the form $\prod_{i\in I}P_i$.

A {\bf marking} of the TG $G_P$ is a pair 
\be{sdffa3343gdsa}(G,\{\beta_V\in Fm\mid V\in G\})\ee
where $G$  is a subset of the set of nodes of $G_P$, 
such that
\bi
\i $G^0_P\in G$, and 
\i $\forall\,V\in G$, if $G_P$
has an edge of the form  $V'\to V$, then $V'\in G$.\ei

Marking \re{sdffa3343gdsa} is said to be
{\bf correct}, if 
\bi\i $G^0_P\models \beta_{G^0_P}$, and
\i $\forall\,S,S'\in\Sigma_P$, if $S\to S'$
and $V^S,V^{S'}\in G$, then the following
implication holds:
$$S\models \beta_{V^S}\;\;\Rightarrow\;\;
S'\models \beta_{V^{S'}}.$$
\ei 

It was noted in section \ref{dsfasfdsagfg} that
for each DP $P$ the graph $G_{P^*}$ 
is a  completion of the graph $G_P$ 
with cyclic edges corresponding to  actions of the 
adversary $P_*$.
Therefore, for each DP $P$, any marking of the TG 
$G_{P}$  
can also be considered as a marking 
of the corresponding TG $G_{P^*}$.

Below, a marking of any TG $G_{P}$ is said to be 
correct, if it is a correct marking (in the sense of the above definition) of the corresponding TG $G_{P^*}$.

\subsubsection{Examples of correct markings of transition graphs}
\label{fergstrg5g4g4664}

In this section we present examples of correct markings 
for TGs from section \ref{dfgdshghdfghdsfsd4}.
The correctness of all the markings listed below
can be justified the theorems from section \ref{sadfaaerg}.

Below we denote nodes of TGs by lists of nodes 
of corresponding SPs.

\bn
\i For TG \re{zfdsas3453y46354y325221}
one of  correct markings  has the form
$$G=\{A^0B^0,A^1B^0,A^1B^1\}$$ and
$$\by
\beta_{A^0B^0}\eam 
{\def\arraystretch{0.5}
\left\{\by 
M_{c_{AB}}=\emptyset\\c_{AB}\,\bot\,  P_*
\\c_{AB}\,\bot\, Channels\ey\right\}},\;
\beta_{A^1B^0}\eam 
{\def\arraystretch{0.5}
\left\{\by 
M_{c_{AB}}=\{x\}\\c_{AB}\,\bot\,   P_*
\\c_{AB}\,\bot\, Channels\ey\right\}}
\\
\beta_{A^1B^1}\eam 
{\def\arraystretch{0.5}
\{
x=y
\}}\ey
$$
\i For TG 
\re{zfdsas3453y46354y3252211327}
one of  correct markings  has the form
$$G=\{A^0B^0,A^1B^0,A^1B^1\}$$ and
$$\by
\beta_{A^0B^0}\eam 
{\def\arraystretch{0.5}
\left\{\by 
k_{AB}^{-1}(M_\circ)=\emptyset\\
k_{AB}\,\bot_{\bf K}\,  P_*
\\k_{AB}\,\bot_{\bf K}\, Channels\ey\right\}},\;
\beta_{A^1B^0}\eam 
{\def\arraystretch{0.5}
\left\{\by 
{k_{AB}}^{-1}(M_{\circ})=\{x\}\\k_{AB}\,\bot_{\bf K}\,  P_*
\\k_{AB}\,\bot_{\bf K}\, Channels\ey\right\}},\\
\beta_{A^1B^1}\eam 
{\def\arraystretch{0.5}
\{
x=y\}}
\ey$$
\i For TG 
\re{zfdsas3453y46354y51}, where actions 
$\alpha_i,\beta_i, \gamma_i\;(i=1,2)$
are defined according to 
\re{sdfgdsfgdsfgw3rt4ge46h5urjhyr},
one of  correct markings  has the form\be{afgaegr3w45y43564ujhbv}
G=\left\{\by
A^0T^0B^0,
A^1T^0B^0,
A^2T^0B^0,
A^1T^1B^0,
A^2T^1B^0,\\
A^1T^2B^0,
A^2T^2B^0,
A^1T^2B^1,
A^2T^2B^1,
A^2T^2B^2
\ey\right\}\ee 
and
\bi\i
$
\beta_{A^0T^0B^0}\eam 
{\def\arraystretch{0.5}
\left\{\by 
M_{c_{AT}}=M_{c_{BT}}=M_{c_{AB}}=
\emptyset\\
\{c_{AT},c_{BT},c_{AB}\}\,\bot\,  P_*
\\\{c_{AT},c_{BT},c_{AB}\}\,\bot\, Channels\ey\right\}}$,
\i $\beta_{A^1T^0B^0}\eam 
{\def\arraystretch{0.5}
\left\{\by 
M_{c_{AT}}=\{c_{AB}\}\\
M_{c_{BT}}=M_{c_{AB}}=
\emptyset\\
\{c_{AT},c_{BT},c_{AB}\}\,\bot\,  P_*
\\\{c_{AT},c_{BT}\}\,\bot\, Channels\ey\right\}}$,
\i $\beta_{A^2T^0B^0}\eam 
{\def\arraystretch{0.5}
\left\{\by 
M_{c_{AT}}=\{c_{AB}\}\\
M_{c_{BT}}=\emptyset\\
M_{c_{AB}}=\{x\}\\
\{c_{AT},c_{BT},c_{AB}\}\,\bot\,  P_*
\\\{c_{AT},c_{BT}\}\,\bot\, Channels\ey\right\}}$,
\i $\beta_{A^1T^1B^0}\eam 
{\def\arraystretch{0.5}
\left\{\by 
u=c_{AB}\\
M_{c_{AT}}=\{c_{AB}\}\\
M_{c_{BT}}=M_{c_{AB}}=
\emptyset\\
\{c_{AT},c_{BT},c_{AB}\}\,\bot\,  P_*
\\\{c_{AT},c_{BT}\}\,\bot\, Channels\ey\right\}}$,
\i $\beta_{A^2T^1B^0}\eam 
{\def\arraystretch{0.5}
\left\{\by 
u=c_{AB}\\
M_{c_{AT}}=\{c_{AB}\}\\
M_{c_{BT}}=\emptyset\\
M_{c_{AB}}=\{x\}\\
\{c_{AT},c_{BT},c_{AB}\}\,\bot\,  P_*
\\\{c_{AT},c_{BT}\}\,\bot\, Channels\ey\right\}}$,
\i $\beta_{A^1T^2B^0}\eam 
{\def\arraystretch{0.5}
\left\{\by 
u=c_{AB}\\
M_{c_{AT}}=\{c_{AB}\}\\
M_{c_{BT}}=\{u\}\\
M_{c_{AB}}=
\emptyset\\
\{c_{AT},c_{BT},c_{AB}\}\,\bot\,  P_*
\\\{c_{AT},c_{BT}\}\,\bot\, Channels\ey\right\}}$,
\i $\beta_{A^2T^2B^0}\eam 
{\def\arraystretch{0.5}
\left\{\by 
u=c_{AB}\\
M_{c_{AT}}=\{c_{AB}\}\\
M_{c_{BT}}=\{u\}\\
M_{c_{AB}}=\{x\}\\
\{c_{AT},c_{BT},c_{AB}\}\,\bot\,  P_*
\\\{c_{AT},c_{BT}\}\,\bot\, Channels\ey\right\}}$,
\i $\beta_{A^1T^2B^1}\eam 
{\def\arraystretch{0.5}
\left\{\by 
u=c_{AB}\\
v=u\\
M_{c_{AT}}=\{c_{AB}\}\\
M_{c_{BT}}=\{u\}\\
M_{c_{AB}}=
\emptyset\\
\{c_{AT},c_{BT},c_{AB}\}\,\bot\,  P_*
\\\{c_{AT},c_{BT}\}\,\bot\, Channels\ey\right\}}$,
\i $\beta_{A^2T^2B^1}\eam 
{\def\arraystretch{0.5}
\left\{\by 
u=c_{AB}\\
v=u\\
M_{c_{AT}}=\{c_{AB}\}\\
M_{c_{BT}}=\{u\}\\
M_{c_{AB}}=
\{x\}\\
\{c_{AT},c_{BT},c_{AB}\}\,\bot\,  P_*
\\\{c_{AT},c_{BT}\}\,\bot\, Channels\ey\right\}}$,
\i $\beta_{A^2T^2B^2}\eam\{y=x\}$
\ei

\i For TG 
\re{zfdsas3453y46354y51}, where actions 
$\alpha_i,\beta_i, \gamma_i\;(i=1,2)$
are defined according to 
\re{sdfgdsfgdsfgw3rt4ge46h5urjhyr1},
one of  correct markings  has the form:
\bi\i
$G$ has the same form, as in 
\re{afgaegr3w45y43564ujhbv},
and
\i
\bi\i
$
\beta_{A^0T^0B^0}\eam 
{\def\arraystretch{0.5}
\left\{\by 
k_{AT}^{-1}(M_\circ)=k_{BT}^{-1}(M_\circ)=
k_{AB}^{-1}(M_\circ)=
\emptyset\\
\{k_{AT},k_{BT},k_{AB}\}\,\bot_{\bf K}\,  P_*
\\\{k_{AT},k_{BT},k_{AB}\}\,\bot_{\bf K}\, Channels\ey\right\}}$,
\i $\beta_{A^1T^0B^0}\eam 
{\def\arraystretch{0.5}
\left\{\by 
k_{AT}^{-1}(M_\circ)=\{k_{AB}\}\\
k_{BT}^{-1}(M_\circ)=k_{AB}^{-1}(M_\circ)=
\emptyset\\
\{k_{AT},k_{BT},k_{AB}\}\,\bot_{\bf K}\,  P_*
\\\{k_{AT},k_{BT},k_{AB}\}\,\bot_{\bf K}\, Channels\ey\right\}}$,
\i $\beta_{A^2T^0B^0}\eam 
{\def\arraystretch{0.5}
\left\{\by 
k_{AT}^{-1}(M_\circ)=\{k_{AB}\}\\
k_{BT}^{-1}(M_\circ)=\emptyset\\
k_{AB}^{-1}(M_\circ)=\{x\}\\
\{k_{AT},k_{BT},k_{AB}\}\,\bot_{\bf K}\,  P_*
\\\{k_{AT},k_{BT},k_{AB}\}\,\bot_{\bf K}\, Channels\ey\right\}}$,
\i $\beta_{A^1T^1B^0}\eam 
{\def\arraystretch{0.5}
\left\{\by 
u=k_{AB}\\
k_{AT}^{-1}(M_\circ)=\{k_{AB}\}\\
k_{BT}^{-1}(M_\circ)=
k_{AB}^{-1}(M_\circ)=
\emptyset\\
\{k_{AT},k_{BT},k_{AB}\}\,\bot_{\bf K}\,  P_*
\\\{k_{AT},k_{BT},k_{AB}\}\,\bot_{\bf K}\, Channels\ey\right\}}$,
\i $\beta_{A^2T^1B^0}\eam 
{\def\arraystretch{0.5}
\left\{\by 
u=k_{AB}\\
k_{AT}^{-1}(M_\circ)=\{k_{AB}\}\\
k_{BT}^{-1}(M_\circ)=\emptyset\\
k_{AB}^{-1}(M_\circ)=\{x\}\\
\{k_{AT},k_{BT},k_{AB}\}\,\bot_{\bf K}\,  P_*
\\\{k_{AT},k_{BT},k_{AB}\}\,\bot_{\bf K}\, Channels\ey\right\}}$,
\i $\beta_{A^1T^2B^0}\eam 
{\def\arraystretch{0.5}
\left\{\by 
u=k_{AB}\\
k_{AT}^{-1}(M_\circ)=\{k_{AB}\}\\
k_{BT}^{-1}(M_\circ)=\{u\}\\
k_{AB}^{-1}(M_\circ)=
\emptyset\\
\{k_{AT},k_{BT},k_{AB}\}\,\bot_{\bf K}\,  P_*
\\\{k_{AT},k_{BT},k_{AB}\}\,\bot_{\bf K}\, Channels\ey\right\}}$,
\i $\beta_{A^2T^2B^0}\eam 
{\def\arraystretch{0.5}
\left\{\by 
u=k_{AB}\\
k_{AT}^{-1}(M_\circ)=\{k_{AB}\}\\
k_{BT}^{-1}(M_\circ)=\{u\}\\
k_{AB}^{-1}(M_\circ)=\{x\}\\
\{k_{AT},k_{BT},k_{AB}\}\,\bot_{\bf K}\,  P_*
\\\{k_{AT},k_{BT},k_{AB}\}\,\bot_{\bf K}\, Channels\ey\right\}}$,
\i $\beta_{A^1T^2B^1}\eam 
{\def\arraystretch{0.5}
\left\{\by 
u=k_{AB}\\
v=u\\
k_{AT}^{-1}(M_\circ)=\{k_{AB}\}\\
k_{BT}^{-1}(M_\circ)=\{u\}\\
k_{AB}^{-1}(M_\circ)=
\emptyset\\
\{k_{AT},k_{BT},k_{AB}\}\,\bot_{\bf K}\,  P_*
\\\{k_{AT},k_{BT},k_{AB}\}\,\bot_{\bf K}\, Channels\ey\right\}}$,
\i $\beta_{A^2T^2B^1}\eam 
{\def\arraystretch{0.5}
\left\{\by 
u=k_{AB}\\
v=u\\
k_{AT}^{-1}(M_\circ)=\{k_{AB}\}\\
k_{BT}^{-1}(M_\circ)=\{u\}\\
k_{AB}^{-1}(M_\circ)=
\{x\}\\
\{k_{AT},k_{BT},k_{AB}\}\,\bot_{\bf K}\,  P_*
\\\{k_{AT},k_{BT},k_{AB}\}\,\bot_{\bf K}\, Channels\ey\right\}}$,
\i $\beta_{A^2T^2B^2}\eam\{y=x\}$.
\ei
\ei
\en

\subsubsection{Application of markings of 
transition graphs in the problems of  verification of 
distributed processes}

An 
{\bf execution} of a DP $P$ is a 
sequence $S_0,\ldots, S_n$ of states from 
$\Sigma_P$, such as $S_0= \odot$, and 
either $n=0$, or 
$\forall\,i=0,\ldots, n-1\quad S_{i}\to S_{i+1}$.

It is not difficult to see that each such sequence $ S_0, \ldots, S_n $ corresponds to a path
$G^0_P=V^{S_0}\to 
\ldots\to V^{S_n}$ in TG $G_P$.

Some correctness properties of DPs have the following form: 
in each state $S$ of an arbitrary execution of a DP $P$, the following implication holds:
\be{dsfgdsgdsgewrwt3w3}
S\models \beta\;\Rightarrow\;
S\models \beta',\mbox{
 where $\beta,\beta'\in Fm$ are given formulas.}\ee
 
For example, for the DP $P^*$, where $P$ 
is any of the DPs presented in section
 \ref{pkergnmf43564654632},
  one of the correctness properties has the following form:
for arbitrary execution $S_0, \ldots, S_n$ of this DP, 
\bi \i if \bi \i $S_n\models (v^{S_n}_B=B^1)$ 
(for first and second DPs in section
 \ref{pkergnmf43564654632}), or \i 
 $S_n\models (v^{S_n}_B=B^2)$ 
 (for third and fourth 
 DPs in section \ref {pkergnmf43564654632}), \ei
i.e. if $B$ executed  an action of receiving 
the message sent by $A$ and wrote the received message in 
variable $y$, \i then $S_n \models (x = y)$, i.e. the received message is the same as the message $x$ that $A$ sent $B$.
\ei

Properties of the form \re{dsfgdsgdsgewrwt3w3} 
can be verified using  a marking of TG $G_P$ 
of an analyzed DP $P$ as follows: 
\bi
\i a correct marking
$(G,\{\beta_V\in Fm\mid V\in G\})$ of $G_P$
 is being built, and
\i for each node 
$V \in G$, such that 
$\beta_V$ implies $\beta$,
the implication  $\beta_V\Rightarrow \beta'$
is being checked.
\ei

To check the above statements, there is no need 
to fully build the TG $G_P$ of the analyzed DP $P$. 
It is convenient to build the TG together with 
the construction of its marking as follows: if a formula 
$\beta_V$ is built to mark the node 
$V$ of $G_P$, and $\beta_V$ 
implies an unrealizability of some edge outgoing from $V$, then this edge is discarded.
For example, this can happen if 
\bi \i a label of an edge outgoing from $V$ 
is of the form $(c?\hat y)_B$, and \i $\beta_V$ contains 
the conjunctive term $M_c = \emptyset$. \ei

As a result of such a construction with discarding unrealizable edges, a fragment of the TG $G_P$ will be obtained.
We shall call such fragment a {\bf reduced TG}. 

It is not difficult to see that the reduced TG preserves 
all the properties of the TG $G_P$. In particular, for the solution of the verification problem described above for a property of the form \re{dsfgdsgdsgewrwt3w3}, the corresponding reduced TG can be used instead of the TG $G_P$.

\subsubsection{Reduction of transition graphs}

\bn\i The
edge
$A^0B^0\ral{(\bar c_{AB}? \hat y)_B}A^0B^1$
in TG \re{zfdsas3453y46354y325221}
is unrealizable.

The reduced TG \re{zfdsas3453y46354y325221}
has the form 

\be{zfdsas3453y4346354y325221}
\by
\begin{picture}(100,10)
\put(-100,0){\oval(34,20)}
\put(-100,0){\oval(38,24)}
\put(-100,0){\makebox(0,0)[c]{${ A^0B^0}$}}

\put(0,0){\oval(34,20)}
\put(0,0){\makebox(0,0)[c]{${A^1B^0}$}}

\put(100,0){\oval(34,20)}
\put(100,0){\makebox(0,0)[c]{${A^1B^1}$}}

\put(-81,0){\vector(1,0){64}}
\put(17,0){\vector(1,0){66}}

\put(50,2){\makebox(0,0)[b]{
$(\bar c_{AB}?\hat y)_B$
}}
\put(-50,2){\makebox(0,0)[b]{
$(\bar c_{AB}!x)_A$
}}
\put(117,4){\vector(3,1){20}}
\put(117,-4){\vector(3,-1){20}}
\put(130,0){\makebox(0,0){
$\ldots$}}
\end{picture}
\ey
\ee

\i

The edge
$A^0B^0\ral{?\bar k_{AB}(\hat y)}A^0B^1$
in TG \re{zfdsas3453y46354y3252211327}
is unrealizable.

The reduced TG \re{zfdsas3453y46354y3252211327}
has the form 
\be{zfdsas3453y43463522354y325221}
\by
\begin{picture}(100,15)
\put(-100,0){\oval(34,20)}
\put(-100,0){\oval(38,24)}
\put(-100,0){\makebox(0,0)[c]{${ A^0B^0}$}}

\put(0,0){\oval(34,20)}
\put(0,0){\makebox(0,0)[c]{${A^1B^0}$}}

\put(100,0){\oval(34,20)}
\put(100,0){\makebox(0,0)[c]{${A^1B^1}$}}

\put(-81,0){\vector(1,0){64}}
\put(17,0){\vector(1,0){66}}

\put(50,2){\makebox(0,0)[b]{
$?\bar k_{AB}(\hat y)$
}}
\put(-50,2){\makebox(0,0)[b]{
$!\bar k_{AB}(x)$
}}
\put(117,4){\vector(3,1){20}}
\put(117,-4){\vector(3,-1){20}}
\put(130,0){\makebox(0,0){
$\ldots$}}

\end{picture}
\ey
\ee

\i

The following edges in TG \re{zfdsas3453y46354y51}
(where actions 
$\alpha_i,\beta_i, \gamma_i\;(i=1,2)$
are defined according to
\re{sdfgdsfgdsfgw3rt4ge46h5urjhyr}
or according to \re{sdfgdsfgdsfgw3rt4ge46h5urjhyr1})
are unrealizable:
$$\by
A^0T^0B^0\ra{\beta_1} A^0T^0B^1\\
A^0T^0B^0\ra{\gamma_1}A^0T^1B^0\\
A^1T^0B^0\ra{\beta_1} A^1T^0B^1\\
A^1T^1B^0\ra{\beta_1}A^1T^1B^1\\
A^1T^2B^1\ra{\beta_2}A^1T^2B^2\\
A^2T^0B^0\ra{\beta_1}A^2T^0B^1\\
A^2T^1B^0\ra{\beta_1}A^2T^1B^1
\ey$$

The reduced TG \re{zfdsas3453y46354y51} has the form 

{\def\arraystretch{1}
{\small
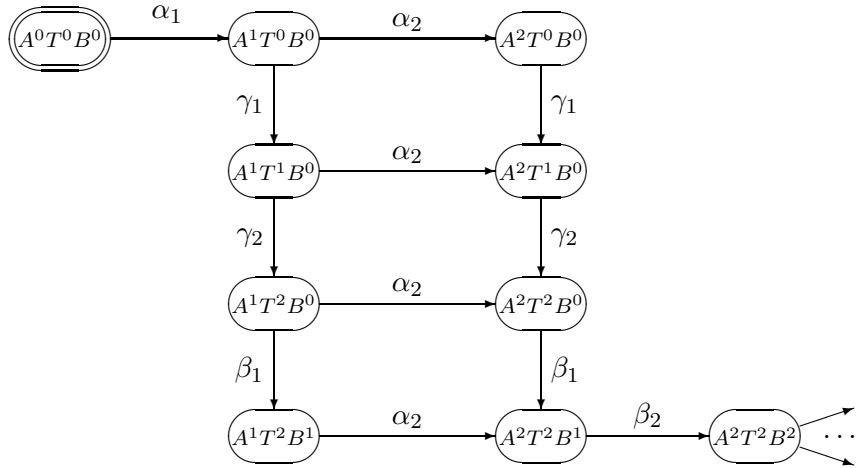
$$\by
\begin{picture}(50,170)

\put(147,4){\vector(3,1){20}}
\put(147,-4){\vector(3,-1){20}}
\put(160,0){\makebox(0,0){
$\ldots$}}

\put(-130,150){\oval(34,20)}
\put(-130,150){\oval(38,24)}
\put(-130,150){\makebox(0,0)[c]{${\scriptstyle A^0T^0B^0}$}}

\put(-50,150){\oval(34,20)}
\put(-50,150){\makebox(0,0)[c]{${\scriptstyle A^1T^0B^0}$}}

\put(-50,100){\oval(34,20)}
\put(-50,100){\makebox(0,0)[c]{${\scriptstyle A^1T^1B^0}$}}

\put(-50,50){\oval(34,20)}
\put(-50,50){\makebox(0,0)[c]{${\scriptstyle A^1T^2B^0}$}}

\put(50,150){\oval(34,20)}
\put(50,150){\makebox(0,0)[c]{${\scriptstyle A^2T^0B^0}$}}

\put(50,100){\oval(34,20)}
\put(50,100){\makebox(0,0)[c]{${\scriptstyle A^2T^1B^0}$}}

\put(50,50){\oval(34,20)}
\put(50,50){\makebox(0,0)[c]{${\scriptstyle A^2T^2B^0}$}}

\put(-50,125){\makebox(0,0)[r]{
$\gamma_1$
}}

\put(50,125){\makebox(0,0)[l]{
$\gamma_1$
}}

\put(-50,75){\makebox(0,0)[r]{
$\gamma_2$
}}

\put(50,75){\makebox(0,0)[l]{
$\gamma_2$
}}

\put(-90,156){\makebox(0,0)[b]{
$\alpha_1$
}}

\put(0,153){\makebox(0,0)[b]{
$\alpha_2$
}}

\put(0,103){\makebox(0,0)[b]{
$\alpha_2$
}}

\put(0,53){\makebox(0,0)[b]{
$\alpha_2$
}}

\put(-50,140){\vector(0,-1){30}}
\put(50,140){\vector(0,-1){30}}

\put(-50,90){\vector(0,-1){30}}
\put(50,90){\vector(0,-1){30}}

\put(-111,150){\vector(1,0){44}}

\put(-33,150){\vector(1,0){66}}
\put(-33,100){\vector(1,0){66}}
\put(-33,50){\vector(1,0){66}}

\put(-50,0){\oval(34,20)}
\put(-50,0){\makebox(0,0)[c]{${\scriptstyle A^1T^2B^1}$}}

\put(50,0){\oval(34,20)}
\put(50,0){\makebox(0,0)[c]{${\scriptstyle A^2T^2B^1}$}}

\put(0,3){\makebox(0,0)[b]{
$\alpha_2$
}}

\put(-33,0){\vector(1,0){66}}

\put(130,0){\oval(34,20)}
\put(130,0){\makebox(0,0)[c]{${\scriptstyle A^2T^2B^2}$}}

\put(-50,40){\vector(0,-1){30}}
\put(50,40){\vector(0,-1){30}}
\put(67,0){\vector(1,0){46}}

\put(-50,25){\makebox(0,0)[r]{
$\beta_1$
}}

\put(50,25){\makebox(0,0)[l]{
$\beta_1$
}}
\put(90,3){\makebox(0,0)[b]{
$\beta_2$
}}

\end{picture}
\ey
$$
 }
}\vspace{0mm}
\en

Note that in all reduced TGs there is a single node
$S$ such that \bi \i $S\models (v^{S}_B=B^1)$ (for first and second DPs in section \ref{pkergnmf43564654632}), and
\i $S\models (v^{S}_B=B^2)$ 
(for third and fourth DPs in section
\ref{pkergnmf43564654632}). \ei

There are correct markings of these reduced TGs presented in
section \ref {fergstrg5g4g4664} 
such that $\beta_{V^S}=(x=y)$.
As stated above, this statement is a justification of the 
property $S\models (x=y)$

Thus, by building a suitable marking, we verified the following property of all four considered DPs: if $B$ executed 
the action of receiving the message sent by $A$ and wrote the received message in the variable $y$,
then the received message is the same 
as the message  that $A$ sent $B$.

\section{An example of a cryptographic protocol verification}

\subsection{Description of a cryptographic protocoll}

In this section we consider an example of a 
cryptographic protocol 
for transmitting encrypted messages 
between multiple  agents
through the open channel $\circ$.
The participants of this protocol are \bi\i
agents from the set
 ${\bf A}=\{A_1,\ldots, A_n\}$, and
\i a trusted intermediary $T$, 
with  use of which agents from 
the set ${\bf A}$ send messages to each other. \ei

Each agent $A_i\in {\bf A}$ 
uses the key $k_{A_iT}$ to communicate with $T$,
which is known only to agent $A_i$ and $T$.
A session of a transmission of an encrypted message $x$ from agent $A_i \in {\bf A}$ to agent $A_j \in {\bf A}$ is a modification of the Wide Mouth Frog protocol.
This session is denoted by the notation $A_i \ra{x} A_j$, 
and is consisting of the following actions:
\bi \i an exchange messages between $A_i$ and $T$, 
resulting in  $T$ finds out
 \bi \i the sender's name $A_i$, 
 the recipient's  name $A_j$, and \i
the key $k_{A_iA_j}$,  
on which the message $x$ from $A_i$ to $A_j$ 
will be encrypted,
\ei
\i an exchange messages between $T$ and $A_j$, 
resulting in $A_j$ finds out
\bi \i the sender's name $A_i$ of the 
message that $A_j $ will receive from $A_i$, 
\i the key $k_{A_iA_j}$ on which this message will be encrypted, \ei
\i sending the encrypted message
$ k_{A_iA_j} (x, \ldots) $ from $ A_i $ to $ A_j $.
\ei

This session is represented by the following scheme:
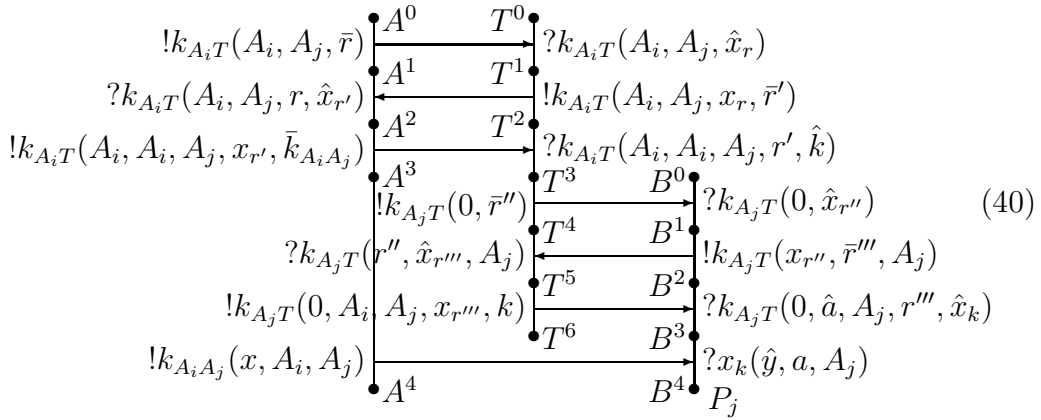
\be{sdfgdsgwer33r333}\hspace{5mm}\by
\begin{picture}(0,145)
\put(-60,140){\circle*{4}}
\put(-57,140){\makebox(0,0)[l]{$A^0$}}
\put(-60,120){\circle*{4}}
\put(-57,120){\makebox(0,0)[l]{$A^1$}}
\put(-60,100){\circle*{4}}
\put(-57,100){\makebox(0,0)[l]{$A^2$}}
\put(-60,80){\circle*{4}}
\put(-57,81){\makebox(0,0)[l]{$A^3$}}
\put(-60,0){\circle*{4}}
\put(-57,0){\makebox(0,0)[l]{$A^4$}}
\put(0,140){\circle*{4}}
\put(-3,140){\makebox(0,0)[r]{$T^0$}}
\put(0,120){\circle*{4}}
\put(-3,120){\makebox(0,0)[r]{$T^1$}}
\put(0,100){\circle*{4}}
\put(-3,100){\makebox(0,0)[r]{$T^2$}}
\put(-60,130){\vector(1,0){60}}
\put(0,110){\vector(-1,0){60}}
\put(-60,90){\vector(1,0){60}}
\put(-60,10){\vector(1,0){120}}
\put(-60,0){\line(0,1){140}}
\put(-63,130){\makebox(0,0)[r]{$!k_{A_iT}(A_i,A_j,\bar r)$}}
\put(3,130){\makebox(0,0)[l]{$?k_{A_iT}(A_i,A_j,\hat x_r)$}}
\put(-63,110){\makebox(0,0)[r]{$?k_{A_iT}(A_i,A_j,r,\hat x_{r'})$}}
\put(3,110){\makebox(0,0)[l]{$!k_{A_iT}(A_i,A_j,x_r,\bar r')$}}
\put(-63,90){\makebox(0,0)[r]{$!k_{A_iT}(A_i,A_i,A_j,x_{r'}, \bar k_{A_iA_j})$}}
\put(3,92){\makebox(0,0)[l]{$?k_{A_iT}(A_i,A_i,A_j, r',\hat k)$}}
\put(-2,68){\makebox(0,0)[r]{$!k_{A_jT}(0,\bar r'')$}}
\put(63,70){\makebox(0,0)[l]{$?k_{A_jT}(0,\hat x_{r''})$}}
\put(-3,50){\makebox(0,0)[r]{$?k_{A_jT}(r'', \hat x_{r'''},A_j)$}}
\put(63,50){\makebox(0,0)[l]{$!k_{A_jT}(x_{r''},\bar r''',A_j)$}}
\put(-3,30){\makebox(0,0)[r]{$!k_{A_jT}(0,A_i,A_j,x_{r'''},
 k)$}}
\put(63,30){\makebox(0,0)[l]{$?k_{A_jT}(0,
\hat a, A_j,{r'''},\hat x_k)$}}
\put(-63,10){\makebox(0,0)[r]{$
!k_{A_iA_j}(x,A_i,A_j)$}}
\put(63,10){\makebox(0,0)[l]{$?x_k(\hat y, a, A_j)$}}

\put(0,80){\circle*{4}}
\put(3,79){\makebox(0,0)[l]{$T^3$}}
\put(0,60){\circle*{4}}
\put(3,60){\makebox(0,0)[l]{$T^4$}}
\put(0,40){\circle*{4}}
\put(3,40){\makebox(0,0)[l]{$T^5$}}
\put(0,20){\circle*{4}}
\put(3,20){\makebox(0,0)[l]{$T^6$}}
\put(60,80){\circle*{4}}
\put(57,79){\makebox(0,0)[r]{$B^0$}}
\put(60,60){\circle*{4}}
\put(57,60){\makebox(0,0)[r]{$B^1$}}
\put(60,40){\circle*{4}}
\put(57,40){\makebox(0,0)[r]{$B^2$}}
\put(60,20){\circle*{4}}
\put(57,20){\makebox(0,0)[r]{$B^3$}}
\put(60,0){\circle*{4}}
\put(57,0){\makebox(0,0)[r]{$B^4$}}
\put(65,-5){\makebox(0,0)[l]{$P_j$}}
\put(0,70){\vector(1,0){60}}
\put(60,50){\vector(-1,0){60}}
\put(0,30){\vector(1,0){60}}
\put(60,0){\line(0,1){80}}
\put(0,20){\line(0,1){120}}
\end{picture}\ey
\ee

We denote \bi\i by the notations $A_{ij}$,$T_{ij}$ and $B_j$
 the SPs corresponding to the left, middle and right threads
 of this diagram, these SPs describe the work of the sender ($A_i$), a trusted intermediary 
($T$) and the recipient ($A_j$) respectively in this session, and
\i by the symbol $T$ the SP $\sum_{i,j=1}^n T_{ij}$,
which denotes the work of a trusted intermediary
in an arbitrary session of this protocol. \ei

Let a finite set of sessions be given:
\be{fdgdsgaerw3}
A_{i_1}\ra{x_1}A_{j_1},\;\ldots,\;
A_{i_m}\ra{x_m}A_{j_m}.\ee

One of  cryptographic protocols designed to implement this set of sessions is represented by a DP
\be{sdfaasfgasrrrrr}P=
({A_{i_1j_1}}(x_{1}/x),\ldots, 
{A_{i_mj_m}}(x_{m}/x),
T^\infty,
 B^\infty_1,\ldots, B^\infty_n)
\ee

This DP consists of SPs of the following families: $A$, $T$, 
$B_1$, $\ldots$, $B_n$.

$\forall \, i \geq 1 $ we  denote those variables 
of the $i$--th copy of the SP $B_j$ in $ B^\infty_j$, 
which are obtained 
by renaming the corresponding variables of $B_j$, 
by  $x^{(i)}$, 
where $x$ is the corresponding variable of $B_j$.

A property of this protocol that must be verified is the following:
\be{as32dfgagdsfgdgwe343}\by
\mbox{$ \forall \, S \in \Sigma_P $, 
$ \forall \, j = 1, \ldots, n $, 
$ \forall \, i \geq 1 $, if 
$S\models (v^{(i)}_{B_j}=B^4)$,}\\
\mbox{then $M_\circ^S$ 
has a pair of messages of the form
}\\
\mbox{$\bar k_{A_iA_j}(x,A_i,A_j,r)$ and $
\bar k_{A_iT}(A_i,A_j, r)$}\ey
\ee
which means the following: 
a session from \re{fdgdsgaerw3} 
of the form $A_i\ra x A_j$  was executed correctly.

\subsection{Verification of the protocol}

Let $S \in \Sigma_{P}$, 
where $P$ is a DP of the form \re{sdfaasfgasrrrrr}.

Using theorem \ref{doredtuslna32b322e3q} from section 
\ref{sadfaaerg}, it is not so difficult to prove that
\be{sdfgfdssghdsfghsd44}
\forall\,i,j=1,\ldots, n\quad
\{k_{A_iT},k_{A_iA_j}\}\,\bot_{\bf K}\,P_*,\quad
\{k_{A_iT},k_{A_iA_j}\}\,\bot_{\bf K}\,M^S_\circ\ee

Let $\tilde M _{\circ}^S$ be the set of messages 
in $ M^S_\circ$ of the form $k_{A_iT} (\ldots)$ 
and $k_{A_iA_j} (\ldots)$.
Using \re{sdfgdsgwer33r333} and \re{sdfgfdssghdsfghsd44}, it is not so difficult to prove that
every message in $\tilde M _{\circ}^S$ 
has one of the following seven forms:
\be{adsfgadgfdsgsdad1}k_{A_iT}(A_i,A_j,r),\ee
\be{adsfgadgfdsgsdad2}k_{A_iT}(A_i,A_j,r,r'),\ee
\be{adsfgadgfdsgsdad3}k_{A_iT}(A_i,A_i,A_j,r, 
 k_{A_iA_j}),\ee
\be{adsfgadgfdsgsdad4}k_{A_jT}(0,r),\ee
\be{adsfgadgfdsgsdad5}k_{A_jT}(r,r',A_j),\ee
\be{adsfgadgfdsgsdad6}k_{A_jT}(0,A_i,A_j,r, k),\ee
\be{adsfgadgfdsgsdad7}k_{A_iA_j}(x,A_i,A_j,r).\ee

Let \bi\i $\tilde M_{\ref{adsfgadgfdsgsdad1}}^S$, $\ldots$,
$\tilde M_{\ref{adsfgadgfdsgsdad7}}^S$ be subsets of 
$\tilde M_{\circ}^S$, consisting of messages of the form
\re{adsfgadgfdsgsdad1}, $\ldots$, \re{adsfgadgfdsgsdad7}
respectively,\i
$\rho_{\ref{adsfgadgfdsgsdad1},\ref{adsfgadgfdsgsdad2}}$
be a set of pairs of the form
$(\re{adsfgadgfdsgsdad1}, \re{adsfgadgfdsgsdad2})$,
in each of which the third component 
($r$) listed in \re{adsfgadgfdsgsdad1} 
is the same as the third component ($r$) listed in
\re{adsfgadgfdsgsdad2}, 
\i 
$\rho_{\ref{adsfgadgfdsgsdad2},\ref{adsfgadgfdsgsdad3}}$,
$\rho_{\ref{adsfgadgfdsgsdad4},\ref{adsfgadgfdsgsdad5}}$,
$\rho_{\ref{adsfgadgfdsgsdad5},\ref{adsfgadgfdsgsdad6}}$,
be similar sets of pairs of the form
$(\re{adsfgadgfdsgsdad2}, \re{adsfgadgfdsgsdad3})$,
$(\re{adsfgadgfdsgsdad4}, \re{adsfgadgfdsgsdad5})$,
$(\re{adsfgadgfdsgsdad5}, \re{adsfgadgfdsgsdad6})$.
\ei

Define a binary relation $\rho$ on $\tilde M_{\circ}^S $ 
as the least transitive relation containing 
$\rho_{\ref{adsfgadgfdsgsdad1},\ref{adsfgadgfdsgsdad2}}$,
$\rho_{\ref{adsfgadgfdsgsdad2},\ref{adsfgadgfdsgsdad3}}$,
$\rho_{\ref{adsfgadgfdsgsdad4},\ref{adsfgadgfdsgsdad5}}$,
$\rho_{\ref{adsfgadgfdsgsdad5},\ref{adsfgadgfdsgsdad6}}$,
and satisfying the following conditions:
\bi
\i if $\rho$ contains pairs of the form
\be{dfgdsfgsdfser3eefe}
(\re{adsfgadgfdsgsdad1}, 
\re{adsfgadgfdsgsdad3})\quad
\mbox{and}
\quad
(\re{adsfgadgfdsgsdad4}, \re{adsfgadgfdsgsdad6})
\ee
and the last component in  message \re{adsfgadgfdsgsdad3} 
of the first pair is the same as the last component in message
\re{adsfgadgfdsgsdad6} of the second pair, then $\rho$ contains the pair (\re{adsfgadgfdsgsdad3}, \re{adsfgadgfdsgsdad4}) whose components are the corresponding messages from \re{dfgdsfgsdfser3eefe}, and
\i $\rho$ contains each pair of the form
 (\re{adsfgadgfdsgsdad6}, \re{adsfgadgfdsgsdad7}), 
 in which the keys $k$ and $k_{A_iA_j}$ are equal.
 \ei

Below the notations $\exists_1$ and $\exists _{\leq 1}$ 
are read as ``there is only one'' and ``there is at most one'', respectively.

With use of theorem \ref{doredtuslna32b322e3q},
it is not so difficult to prove that
\be{sdfsadgadfga}\left\{\by
\forall\,e\in   \tilde M_{\ref{adsfgadgfdsgsdad3}}^S\;
\exists_1\,e'\in   \tilde M_{\ref{adsfgadgfdsgsdad1}}^S:
(e',e)\in \rho,\\
\forall\,e\in   \tilde M_{\ref{adsfgadgfdsgsdad6}}^S\;
\exists_1\,e'\in   \tilde M_{\ref{adsfgadgfdsgsdad4}}^S:
(e',e)\in \rho,\\
\forall\,e\in   \tilde M_{\ref{adsfgadgfdsgsdad6}}^S\;
\exists_1\,e'\in   \tilde M_{\ref{adsfgadgfdsgsdad1}}^S:
(e',e)\in \rho,\\
\forall\,e\in   \tilde M_{\ref{adsfgadgfdsgsdad7}}^S\;
\exists_{\leq 1}\,e'\in   \tilde M_{\ref{adsfgadgfdsgsdad1}}^S:
(e',e)\in \rho.\ey\right.\ee

\re{sdfsadgadfga} and theorem \ref{doredtuslna32b322e3q} 
imply the following statement 
$\forall \, S \in \Sigma_P$, $ \forall \, i \geq 1 $, if $ S \models (v_{B_j}^{(i)} = B^4) $, then $M_\circ^S$ contains a pair of messages of the form 
\re{as32dfgagdsfgdgwe343}, i.e. the integrity property of the analyzed protocol is true: if agent $A_j$ performed the action of receiving a message sent by agent $A_i$ and wrote the received message to variable $y_{B_j}$, 
then the received message is the same as 
 the message $x$ that $A_i$ sent $A_j$ in the same session.

\section {Conclusion}

In the present work, a new model of cryptographic protocols was built, and examples of its use for solving problems 
of verification of protocol integrity properties are shown.

The objectives for further development of this model and verification methods based on it are the following:
\bn
\i an automation of synthesis of suitable markings in 
transition graphs  of the analyzed protocols,
\i development of the language of specification of properties of cryptographic protocols, which allow to express e.g. 
\bi
\i properties of confidentiality (secrecy) of transmitted messages, i.e. the adversary's inability to extract any new information about the content of messages intercepted by him,
\i matching properties in authentication protocols, or zero knowledge properties,
\i non-traceability properties in electronic payments,
\i properties of correctness of the votes' counting  in  voting protocols, \ei
\i construction of automated synthesis 
methods of cryptographic protocols 
by describing the properties which the
cryptographic protocols  must satisfy, etc.
\en


\begin{thebibliography}{999}

\bibitem{1} Denning D., Sacco G., Timestamps in Key Distribution Protocols,
     Communications of the ACM, Vol. 24, No. 8, (1981) 533-536.
\bibitem{2}Needham R., Schroeder M., Using Encryption for Authentication
	 in large
     networks of computers, Communications of the ACM, 21(12), (1978)
     993-999.
\bibitem{3} Needham R., Schroeder M., Authentication revisited, Operating Systems
     Review, Vol. 21, No. 1, (1987).
     
\bibitem{kerberos}      
Cervesato I., Jaggard A.D., Scedrov A., Tsay J.-K., Walstad C.,
Breaking and fixing public-key Kerberos,
Information and Computation
Volume 206, Issues 2-4, (2008), Pages 402-424.

\bibitem{15} Lowe G., Breaking and Fixing the Needham-Schroeder Public-Key Protocol
     Using FDR, In Proceedings of TACAS, (1996) 147-166, Springer Verlag.     


\bibitem{kerbprot}
Kerberos: The Network Authentication Protocol. 
MIT Kerberos. 10 September 2015. Retrieved 31 October 2015.\\
\verb'http://web.mit.edu/kerberos/'

\bibitem{5} Burrows M., Abadi M., Needham R., 
A Logic of Authentication. In ACM
     Transactions on Computer Systems, 8(1), (1990) 18-36.


\bibitem{strand1}
F. J. Thayer, J. C. Herzog, and J. D. Guttman. 
Strand spaces: Proving security protocols correct. 
Journal of Computer Security, 7(2/3):191–230, 1999. 

\bibitem{strand2}
J. D. Guttman and F. J. Thayer. 
Authentication Tests and the Structure of Bundles. 
Theoretical Computer Science, June, 2002. 

\bibitem{strandlast}
Joshua D. Guttman.
 State and Progress in Strand Spaces: Proving Fair Exchange. Journal of Automated Reasoning, 48(2): 159–195, 2012. 



\bibitem{hartog}
M. Abadi. Security Protocols and Their Properties. 
In NATO Science Series: Volume for the 20th International Summer School on Foundations of Secure Computation, pp. 39-60, Marktoberdorf, Germany, 1999.

\bibitem{hartog1}
M. Abadi and B. Blanchet. 
Secrecy Types for Asymmetric Communication. 
In Conference on Foundations of Software Science and Computation Structures (FOSSACS), LNCS 2030, pp. 25-41, 2001.

\bibitem{hartog2}
M. Abadi and R. Needham. 
Prudent Engineering Practice for Cryptographic Protocols. 
In IEEE Transactions on Software Engineering, 22(1):6-15, 1996.

\bibitem{hartog3}
M. Abadi and B. Blanchet. 
Analyzing Security Protocols with Secrecy Types and Logic Programs. 
In Journal of the ACM, 52(1), pp. 102-146, 2005.

\bibitem{hartog4}
B. Blanchet. 
An Efficient Cryptographic Protocol Verifier Based on Prolog Rules. In 14th IEEE Computer Security Foundations Workshop (CSFW), pp. 82-96, 2001.

\bibitem{hartog5}
L. C. Paulson. 
Inductive Analysis of the Internet Protocol TLS. 
In ACM Trans. on Information and System Security, 2(3), pp. 332-351, 1999.

\bibitem{hartog6}
J. Zhou, D. Gollmann 
A Fair Non-repudiation Protocol. 
In IEEE Symposium on Research on Security and Privacy, pp. 55-61, 1996.

\bibitem{hartog7}
M. Abadi, N. Glew, B. Horne, B. Pinkas 
Certified E-mail with a Light On-line Trusted Third Party: Design and Implementation. 
In 11th Int. World Wid Web Conference, pp. 387-396, 2002.

\bibitem{hartog8}
M. Abadi, B. Blanchet 
Computer-assisted Verification for Certified E-mail. 
In Science of Computer Programming, 58(1-2):3-27, 2005.

\bibitem{hartog9}
M. Abadi. 
Secrecy by Typing in Security Protocols. 
In Journal of the ACM, 46(5), pp. 749-786, 1999.


\bibitem{hartog10}
S. Kremer, M. Ryan. 
Analysis of an Electronic Voting Protocol in the Applied Pi Calculus. 
In 14th European Symposium on Programming (ESOP), pp. 186-200, 2005.


\bibitem{hartog11}
B. Blanchet. 
Automatic Proof of Strong Secrecy for Security Protocols. 
In IEEE Symposium on Security and Privacy, pp. 86-100, 2004.

\bibitem{hartog12}
B. Blanchet, M. Abadi, C. Fournet. 
Automated Verification of Selected Equivalences for Security Protocols. 
In 20th IEEE Symposium on Logic in Computer Science (LICS), pp. 331-340, 2005.


\bibitem{hartog13}
M. Abadi, P. Rogaway. 
Reconciling Two Views of Cryptography (The Computational Soundness of Formal Encryption). 
In IFIP International Conference on Theoretical Computer Science (IFIP TCS), 2000.

\bibitem{hartog14}
W. Aiello, S. Bellovin, M. Blaze, R. Canetti, J. Ioannidis, A. Keromytis, O. Reingold. 
Just Fast Keying: Key Agreement in a Hostile Internet. 
In ACM Transactions on Information and System Security, 7(2):242-273, 2004.

\bibitem{hartog15}
M. Abadi, B. Blanchet, C. Fournet. Just Fast Keying in the Pi Calculus. 
In ACM Transactions on Information and System Security, 10(3), 2007.

\bibitem{hartoglast}
A. Gordon and A. Jeffrey. 
Authenticity Authenticity by Typing for Security Protocols. 
In Journal of Computer Security, 11(4), pp. 451-521, 2003.


\bibitem{Duncan}
Duncan, Richard. 
An Overview of Different Authentication Methods and Protocols. 
SANS Institute. Retrieved 31 October 2015

\bibitem{Security}
Proceedings of 
Joint Workshop on Foundations of Computer Security and Automated Reasoning for Security Protocol Analysis (FCS-ARSPA ’06)
Information and Computation
Volume 206, Issue 2, (2008).

\bibitem{verc}
 Veronique Cortier, Steve Kremer.
Formal Models and Techniques for Analyzing Security Protocols.
Now Publishers Inc.,
Hanover, United States
(2014).

\bibitem{7} Syverson P., van Oorschot P.C., On Unifying some Cryptographic
     Protocol Logics, Proceedings of the 1994 IEEE Computer Security
     Foundations Workshop VII, (1994) 14-29, IEEE Computer Society Press.
\bibitem{11} Syverson P., Meadows C., A Logical Language for Specifying
     Cryptographic Protocol Requirements, Proceedings of the 1993 IEEE
	 Computer Security Symposium on Security and Privacy, (1993) 165-177,
     IEEE Computer Society Press.
\bibitem{24} Paulson L., Proving Properties of Security Protocols by Induction,
     Proceedings of the IEEE Computer Security Foundations Workshop X,
     (1997) 70-83, IEEE Computer Society Press.
\bibitem{27} Brackin S., A State-Based HOL Theory of Protocol Failure, (1997), ATR
	 98007, Arca Systems, Inc.,
	 http://www.arca.com/paper.htm.

\bibitem{rsm}
Mark D. Ryan and Ben Smyth, Applied pi calculus, 
in: Formal Models and Techniques for Analyzing Security Protocols, Edited by Veronique Cortier,  2011 IOS Press, p. 112-142.

     \bibitem{82} Abadi M., Gordon A., A Calculus for Cryptographic Protocols: The Spi
     Calculus, Proceedings of the Fourth ACM Conference on Computers and
     Communications Security, (1997) 36-47, ACM Press.


\bibitem{abf}
M. Abadi, B. Blanchet, C. Fournet. The Applied Pi Calculus: Mobile Values, New
Names, and Secure Communication. [Research Report] ArXiv. 2016, pp.110. \verb'hal-01423924',
\verb'https://arxiv.org/abs/1609.03003'




\end{thebibliography}
\end{document}